\documentclass[preprint,12pt]{aastex} 
\usepackage{epsfig}

\newcommand{\Lsun}{\mbox{$L_{\odot}$}}
\newcommand{\Msun}{\mbox{$M_{\odot}$}}
\newcommand{\Rsun}{\mbox{$R_{\odot}$}}
\newcommand{\Lstar}{\mbox{$L_*$}}

\newcommand{\Tel}{\mbox{$T_{\rm e}$}}
\newcommand{\nel}{\mbox{$n_{\rm e}$}}
\newcommand{\etal}{\mbox{\it et~al.}}
\newcommand{\Msunyr}{\mbox{$M_{\odot}$ yr$^{-1}$}}
\newcommand{\kms}{\mbox{km~s$^{-1}$}}

\newcommand{\qmdot}{\mbox{$q_{\Delta M}$}}
\newcommand{\tauKH}{\mbox{$\tau_{\rm KH}$}}
\newcommand{\taums}{\mbox{$\tau_{\rm MS}$}}
\newcommand{\taumix}{\mbox{$\tau_{\rm mix}$}}
\newcommand{\tmix}{\mbox{$\tau_{\rm mix}$}}
\newcommand{\taumixratio}{\mbox{$\tau_{\rm mix} / \tau_{\rm MS}$}}

\newcommand{\Mw}{\mbox{$\dot{M}_{\rm w}$}}

\shorttitle{Nebular abundances and the evolution of LBVs}

\begin{document}

\title{ Chemical composition  and origin of nebulae around
  Luminous Blue Variables}

\author{ Henny J.G.L.M. Lamers \altaffilmark{1,2}, 
         Antonella Nota \altaffilmark{3,4},
         Nino Panagia \altaffilmark{3,4}
         Linda J. Smith \altaffilmark{5}
    \and Norbert Langer \altaffilmark{1}}

\altaffiltext{1} {Astronomical Institute, Princetonplein 5, NL-3584 CC Utrecht,
                  The Netherlands;  lamers@astro.uu.nl, langer@astro.uu.nl}
\altaffiltext{2} {SRON Laboratory for Space Research, Sorbonnelaan 2,
                  NL-3584 CA, Utrecht, The Netherlands}
\altaffiltext{3} {Space Telescope Science Institute, 3700 San Martin Drive,
                  Baltimore, MD 21218, USA; nota@stsci.edu, panagia@stsci.edu} 
\altaffiltext{4} {Affiliated with the Astrophysics Division, Space Science 
                  Department of the European Space Agency}
\altaffiltext{5} {University College London, Gower Street, London, UK;
                  ljs@star.ucl.ac.uk}
                       
%

\begin{abstract}
  We use the analysis of the heavy element  abundances (C, N, O, S) in
  circumstellar nebulae around Luminous Blue Variables 
  to infer the evolutionary phase in which the
  material has been ejected. We concentrate on four aspects.\\
  (1) We discuss the different effects that may have changed the gas 
  composition of the nebula since it was ejected: mixing with the
  swept up gas from the wind-blown bubble, mixing with the gas from the
  faster wind of the central star, and depletion by CO and dust.\\
  (2) We calculate the expected  abundance changes at the stellar surface
  due to envelope convection in the red supergiant phase. We show that 
  this depends strongly on the total amount of mass that was lost 
  prior to the onset of the envelope convection. {\it If} the observed LBV
  nebulae are ejected during the red supergiant phase, the abundances
  of the LBV nebulae require a significantly smaller amount of mass
  to be lost than assumed 
  in the evolutionary calculations of Meynet et al. (1994).\\
  (3) We calculate the changes in the surface composition during
  the main sequence phase by rotation induced mixing. 
  {\it If} the nebulae are ejected at the end of the main
  sequence phase, the abundances
  in LBV nebulae are compatible with mixing times between $5 \times
  10^6$ and $1 \times 10^7$ years. These values are reasonable,
  considering the high rotational velocities of main sequence O-stars.
  The existence of ON stars supports this scenario.\\
  (4) The predicted He/H ratio in the nebulae, derived from the 
  observed N/O ratios, are significantly smaller than the current observed
  photospheric values of their central  stars. This indicates
  that either (a) the nebula was ejected from a star that had an
  {\it abundance gradient} in its envelope, or (b) that
  fast mixing on a time scale of $10^4$ years must have occurred in the
  stars {\it immediately after} the nebula was ejected.\\
  Combining various arguments we show that the LBV nebulae are
  ejected during the blue 
  supergiants phase and that the stars have {\it not}
  gone through a red supergiant phase. The chemical enhancements are
  due to rotation induced mixing, and the ejection is possibly 
  triggered by near-critical rotation. 
  During the ejection, the outflow was
  optically thick, which resulted in a large effective
  radius and a low effective temperature. This explains why the
  observed properties of the dust around LBVs  closely 
  resemble the properties of dust  formed around red supergiants.


\end{abstract}



\section{Introduction}
\label{sec:intro}

Luminous Blue Variables (LBVs) are very luminous  stars with log
$L/\Lsun \simeq$ 5.0 to 6.3 and  variable spectral types between about
O9 and A (Humphreys \& Davidson, 1994).  In the HR diagram, they are
located in close proximity to the  observed luminosity upper limit for
very massive stars, the Humphreys-Davidson (HD) limit. This upper limit
suggests that stars above log $L/\Lsun \simeq 5.8$ do not evolve into
red supergiants (RSG) but that they are prevented from doing so by very
severe mass loss in the previous phases.  

From an observational
point of view, LBVs are characterised by extreme instability, violent
eruptions - with visual brightness increases of 3 magnitudes or more -
and high mass loss.  The material ejected (up to several solar
masses) is often  observed in the form of spectacular associated nebulae
(Nota et al. 1995).  In between such dramatic outbursts, LBVs still
lose mass at  high rates -- typically 10$^{-5}$-10$^{-4}$
M$_{\odot}$/yr (e.g. Leitherer  1997).  The properties of LBVs
are reviewed by Humphreys \& Davidson (1994) and in the proceedings of
a  dedicated workshop by Nota \& Lamers (1997).  The ejected nebulae
around LBVs have  been found to have typical diameters of 0.5 to 2 pc,
an expansion velocity between 25 and 140 km s$^{-1}$ , 
and a dynamical age of 5$\times$$10^3$ to
5$\times$$10^4$ years. The exceptions are $\eta$ Car and P Cygni whose ejecta
have higher velocities and much shorter dynamical ages (for a
comprehensive discussion of LBV nebular properties, see review by Nota
\& Clampin, 1997).

LBV's are believed to be the precursors of Wolf-Rayet stars. However,
the evolutionary phase of LBVs is still poorly understood: their
instability and related nebula ejection is believed to occur during
the  post main sequence life of a massive star, but it is not clear
exactly when and how.  When the star is still on the main sequence, the
O star wind creates a wind-blown bubble with a radius of at least 50
pc. This bubble contains the hot, shocked wind from the main sequence
phase. The LBV nebula is then formed within this cavity by rapid mass
loss during the LBV instability phase. Subsequently, in the Wolf-Rayet
phase, the fast and dense
wind  collides with the LBV nebula.  Waters et al (1997, 1998, 1999)
and Voors et al. (2000) have studied the properties of the dust in LBV
nebulae and argued on the basis of the dust properties (composition,
grain size) that the dust is very similar to that of RSGs. They
conclude that the dust in LBV nebulae was ejected under very similar
conditions, most likely when the star was a RSG.

Smith et al. (1998) derived the N and O abundances of several LBV nebulae,
and noticed that the N/O abundance ratios are higher than for 
initial composition gas, but smaller than for gas with CNO-equilibrium
abundances. In fact, they show that the abundances are
quite similar to those predicted for RSGs in the evolutionary models
of the Geneva group (Meynet et al. 1994). Smith et al. 
discuss the problem of reconciling their
nebular N/O ratios with the observed absence of RSG at high
luminosities by postulating a pseudo-RSG phase. They suggest that if
the star does not lose enough mass in the blue phase, than it could
encounter its Eddington limit, develop a convective envelope in
response and become very bloated such that the star will 
appear as a RSG for a brief period.

From a theoretical point of view, two different approaches have
been recently  presented to explain the occurrence of the  LBV instability:
\\1) Langer  et al.  (1994) have proposed  a mechanism based on enhanced
mass loss, dominant towards the end of the core hydrogen
burning evolution.  This high mass loss maintains the star on the blue
side of the HR diagram.  The ignition of the
H-burning shell forces the stellar radius to increase, and the
corresponding redward evolution of the star in the HR diagram results
in a destabilization of the envelope due to the proximity to the
Eddington limit (Lamers \& Fitzpatrick, 1988; Lamers \& Noordhoek
1993). The star experiences  extreme mass loss  (as high as 5 $\times$
10$^{-3}$ M$_\odot$/yr) which  involves the  enriched stellar surface
layers.  This phase
corresponds to the observed LBV instability, and the mass  ejection occurs
therefore when the star is a blue supergiant. Due to the
large mass-loss rates in the LBV phase, the evolution  of the star into
a RSG is avoided.
\\
2) Alternatively, models of Stothers \& Chin (1993, 1996)
predict that the major mass-loss occurs in a single ejection event
during a brief RSG phase. They find that the star rapidly
moves  redward in the HR diagram after core hydrogen exhaustion.  An
instability is then triggered
by the recombination of hydrogen and helium as a result of the envelope
expansion, leading to the rapid loss of the H-rich envelope.  The star
then moves back to the blue and becomes a true LBV.   The mass ejection
occurs while the star is in a brief RSG phase, before the LBV phase.

The problem with the first scenario is that the nature of the
instability is not completely understood.  The challenge to  the second
scenario is that the HD limit shows that no RSG  
exist with $L>6~10^5$ \Lsun\ (Humphreys and Davidson 1979). 
This implies that two of the {\it standard} LBVs,
AG Car in the Galaxy and R\,127 in the LMC, cannot have been RSGs
(except maybe for a short time of $ \lesssim 10^4$ years).

In reality, we still do not know the evolutionary phase at which the
LBV nebulae are ejected. One way to address this problem is to study
the LBV surface chemical composition. However, atmospheric analyses of
heavy element abundances can be very uncertain. Alternatively, we
can study the abundance determinations of the circumstellar  ejected
nebulae, and compare the results with the predicted surface composition
for massive stars during various phases of their evolution, including
the effects of mass loss, convective mixing in the core and in the
envelope and rotation induced mixing.

Some predictions have already been made on what one should expect:
hydrodynamical models of the evolution of the circumstellar environment
of a 60 M$_\odot$ star have been made by Garc\'\i a-Segura et al.
(1996a) as  the star transitions from an O star to an LBV and then to a
WR star. They predict that the LBV nebula will have CNO equilibrium
abundances.  However,  LBV nebular abundance studies of Smith et al.
(1997, 1998)  have shown that the observed N/O ratios are closer to
CN-equilibrium values. This suggests that the nebulae
were created by the ejection of surface layers that consist of a 
mixture of CN-processed gas and original gas.\\

In this paper we study which effects can produce the abundances
that are observed in the LBV nebulae. We consider mixing by
convection of the outer layers in the red supergiant phase,
and by rotation induced mixing during the main sequence phase.\\
(a) In the case of convective mixing, the resulting surface abundance
in the red supergiant phase depends critically 
on the total amount of mass that was lost during the main
sequence phase. If the LBV nebulae are ejected during the
RSG phase, we can determine the amount of mass that was lost
prior to the envelope ejection.\\
(b) In the case of rotation induced mixing,
the surface abundance at the end of the main sequence phase depends
critically on the ratio between the mixing time and the main sequence
lifetime. If the LBV nebulae are ejected in the blue supergiant phase
after the main sequence, we can determine the mixing timescale during
the main sequence phase.\\
We will compare the observed abundances of the LBV nebulae
with both sets of predictions. Combining this with information about the
dynamical age, the mass, the ejection velocity and the 
morphology of the nebulae, we can determine at which evolutionary
stage the LBV nebulae were ejected.

In Sect. \ref{sec:2} we discuss the observed properties of the LBVs. 
In Sect. \ref{sec:3} we describe the effects that may have influenced the
abundances of the nebulae since their ejection.
In Sect. \ref{sec:4} we derive estimates for the mixing in the star,
either convective or rotation-induced,  that must have occurred to
explain the observed N/O and N/S ratios in the nebulae.
Sect. \ref{sec:5} gives a description of the predicted changes in the
surface abundances of massive stars in the red supergiant (RSG) phase
due to the convective envelope, and the way in which the abundances
depend on the mass loss prior to the RSG phase. The results are
compared with the observed abundance ratios of the nebulae.
In Sect. \ref{sec:6} we give a description of 
the changes in the surface composition of the stars due to 
rotation induced mixing during the main sequence phase. 
If the nebulae are ejected in the blue supergiant phase, their abundances
can be used to derive empirical estimates of the mixing time.
In Sect. \ref{sec:7} we apply the results of the previous sections
to determine the evolutionary phase in which the LBV nebulae were ejected.
The results are discussed in Sect. \ref{sec:8}.


\section{The observed properties of LBV nebulae}
\label{sec:2}

\subsection{The nebulae}
\label{sec:2a}

Most  LBVs and related transition objects (eg. Ofpe/WN9 stars) display
associated circumstellar nebulae, which are spatially resolved both 
in ground-based observations  
using coronography and by direct imaging with HST.  
These nebulae  are very similar
in terms of morphological and  physical properties. They are typically
1 parsec in size, with morphologies which are mildly to extremely
bipolar, with the possible exception of the nebula around P Cygni
(Barlow et al. 1994; Nota et al.  2000).  They expand in the
surrounding medium with velocities of the order of 25 -- 140 km
s$^{-1}$.  Their size and expansion velocities identify dynamical
timescales which are of the order of $10^4$ years.  Their
spectra show typical nebular emission lines (H$\alpha$, [NII]
$\lambda\lambda$ 6548, 6583, 5755, [OII] $\lambda\lambda$ 3726, 3729, [SII]
$\lambda\lambda$ 6717, 6731) which can be used to derive the nebular
physical and chemical properties. Densities  derived from the [SII]
$\lambda\lambda$ 6717, 6731 line ratio are generally found to be low
(500 - 1000 cm$^{-3}$) and the temperatures, derived 
from the ratio of the [NII] lines, are in the range 5000 - 10000K. 

The abundances and dynamical stuctures of the LBV nebulae can be used
to study at what phase of the evolution of the central star these
nebulae were
ejected. We have selected four stars for which the nebula has been
studied in detail. These include AG Car and P Cygni in the Galaxy,
and R~127 and S~119 in the LMC. We have omitted the 
well known LBV  $\eta$ Car,
that was studied by Dufour et al. (1997) 
and Ebbets et al. (1997), because of its peculiar nature.
The nebula around R 143 is also omitted, because Smith et al. (1998)
showed that the nebula is not associated with the star.

The properties of the four selected LBVs and their nebulae are summarized
in Table \ref{tbl:parameters}. We list the distance and the luminosity,
but not the radius or effective temperature because these values
are variable for LBVs. We give the photospheric Helium abundance.
The parameters are from compilations by 
Humphreys \& Davidson (1994), Crowther (1997), Nota \& Clampin (1997)
and from Crowther \& Smith (1997).
The Table also shows the parameters of the nebulae,
such as velocity, dynamical age, mean density and temperature, as well
as the nebular abundances. 
The data are from Nota and Clampin (1997),
Smith et al. (1997; 1998) and Schulte-Ladbeck et al. (2000).
Notice that all nebulae have an enhanced 
N-abundance by about 1 dex and a depleted O-abundance by -0.3 to -1.2 dex.
 For AG Carinae, Smith et al. (1997) derived  a nitrogen 
enhancement of a factor 4.5$\pm$1.3 and an oxygen deficiency of a 
factor 15.1$\pm$7.2. Such mild enrichment is typical of material 
which has not reached CNO equilibrium.
For the other two nebulae around R\,127 and S\,119, the
abundance analysis (Smith et al. 1998)  provided very similar results:
 for R\,127, a N
enrichment was found of a factor 10.7$ \pm$2.2, and O depletion of
2.0$\pm$1.0, with N/O = 0.89$\pm$0.40; for S\,119, Smith et al. (1998) were
not able to secure an accurate electron temperature, and found a ratio
N/O = 1.41 to 2.45, similar to the value determined for R\,127 
(Table~1).\footnote{Unless 
otherwise mentioned, all abundance ratios in this paper are
number-ratios, e.g. N/O $\equiv$ n(N)/n(O).}
For the P Cygni nebula, no accurate abundance analysis is currently
available, 
except  for the earlier work by Johnson et al. (1992), where N/S  ratios are
provided at one position in the nebula.   
In Table~1  we also give the number ratios N/H, N/O and N/S, 
which will be compared with
evolutionary calculations. For the N/S ratios we only have upper limits
because they are derived from the N$^+$/S$^+$ ratios with an unknown 
ratio S$^{++}$/S$^{+}$.
Unfortunately we do not have reliable nebular S/H
ratios which might have been used to derive the underabundance of H.

\subsection{The accuracy of the observed abundance ratios}
\label{sec:2b}

The accuracy of the derived abundances can be checked by comparing the
sum of the C, N and O abundances to the initial abundance. Since
the changes in the abundances of these elements are due
to the mixing with products from the CN and the NO-cycles, the
mass ratio of (C+N+O)/(H+He) at the stellar surface 
should remain constant during the evolution
of the star, until the products of He-burning appear at the surface.
If the nebulae had the initial ratio He/H=0.075,
we would expect a number ratio (C+N+O)/H $\simeq 1.6 \times 10^{-3}$ for the
Galactic nebulae and $\simeq 6 \times 10^{-4}$ for the LMC
nebulae. If the He/H ratio in the nebula is about 0.2, as predicted
on the basis of the observed N/O ratios (see later: Tables 2 and 3)
then we expect (C+N+O)/H $\simeq 2 \times 10^{-3}$ for Galactic
nebulae and $8 \times 10^{-4}$ for the LMC nebulae. 
If the He/H ratio in the nebula is 0.40, as it is in the
stellar photospheres, then the mass fraction of H has decreased by
about a factor two and so we would expect ratios of 
(C+N+O)/H $\simeq 3 \times 10^{-3}$ for the
Galactic nebulae and $\simeq 1 \times 10^{-3}$ for the LMC
nebulae. 
We note that the contribution of C is only about 30 percent
for gas with the initial composition, and much less for gas with 
CN-equilibrium composition. So for nebulae enriched in N
we can replace the above mentioned C+N+O abundance
with the N+O abundance to reasonable accuracy.

The data in Table 1 show that the observed number ratio (N+O)/H 
is about a factor two smaller than predicted for R\,127 and S\,119.
For P Cygni the predicted value is within the wide range of the
observed value. For AG Car the observed (N+O)/H ratio is about a factor
five smaller than predicted.
These discrepances are most likely due to errors in the determinations of the
N/H and O/H ratio due to the inhomogeneity of the nebulae.
In a non-homogeneous nebula that contains regions of different
temperatures, either in the form of clumping or in the form of a 
temperature gradient, the strength of H$\beta$ is less affected than
the strength of the forbidden lines, because the latter are 
collisionaly excited from their respective ion ground state,
whereas the former is the product of H recombination.
This results in a systematic overestimate of the electron
temperature and, therefore, in a systematic underestimate in the
derived ratios of metals relative to H
(e.g., Peimbert 1967, Panagia \& Preite Martinez 1975). 
In principle this can be 
tested by determining the S/H ratio, because S is not affected by the
abundance changes inside the star. So the S/H ratio is a good measure
of the changes in the H-content or of the errors in the abundance
determinations due to inhomogeneities in the nebula. Unfortunately we 
only have an upper limit for the N/S ratio of the AG Car nebula, which
does not provide a useful test.

We have calculated the strength of the observed forbidden lines for
simple spherical nebular models with clumping, or with gradients in
temperature or density, for a given input abundance. 
We then used the calculated line ratios to
derive the mean electron density, the mean electron temperature
and the resulting abundances, in the same way as done in the analysis
of the observed spectra. The resulting abundance ratios were then compared
to the input abundance ratios. We found for a large range of models 
and clumping factors, that the thus derived abundance ratios N/H, O/H and S/H 
can differ drastically from the input values, for the reason mentioned
above, but that the abundance ratios of the metals,
N/O and N/S, are very insensitive to clumping and temperature or
density gradients (within about 0.1 dex for a wide range of clumping
factors and gradients). 
For this reason we will concentrate on these
ratios in the rest of this paper.


\section{Does the gaseous nebular abundance reflect the abundance during
ejection?}
\label{sec:3}

The atomic abundances of the gaseous component of the 
LBV nebulae might be different from the 
abundances of the nebula at the time of the ejection
due to several effects, which will be discussed and estimated below:\\ 
(a) a fraction of C and O may have been locked into CO and 
several atomic species may be depleted due to dust, 
in particular O and Si\\
(b) the nebula may have mixed with the gas of the interstellar (IS)  bubble
into which it was ejected \\
(c) the nebula may have mixed with the 
wind from the star after it was ejected. \\

\subsection{Depletion of atomic species by CO and dust}

Infrared studies of LBV nebulae have shown that they contain
significants amounts of CO and dust (e.g. McGregor et al. 1988;
Huts\'emekers 1997). The analysis of
 ISO observations by 
Waters et al. (1997,1998, 1999) and Voors et al. (2000) have shown that
most LBV nebulae contain O-rich dust, with a small contribution
of C-rich PAHs. This dust is mainly in the form of
amorphous silicates, such as  MgFeSiO$_4$  and Fe$_2$SiO$_4$,
plus a minor contribution from
crystalline silicates such as olivines and pyroxenes (Voors et al. 2000). 
 Estimates of the dust mass vary between 
about $10^{-3}$ \Msun\ and $10^{-1}$ \Msun\ (Hutsem\'ekers 1997; 
Lamers et al. 1996; Voors et al. 2000). However, the presence of dust or
molecules does not affect the abundance ratios derived from 
the {\it gaseous} component. This is because the nebular gaseous abundances
are determined from the ionized regions of the nebulae, where no
corrections for the presence of dust, molecules and neutral atoms   
are needed.

\subsection{The interaction of the nebula with the IS bubble or with
  the stellar wind }

The ejected nebula runs into the interstellar bubble that was 
blown during  the main sequence phase. Such an interaction may have 
changed the composition of the nebula from the outside. However,
the density in the bubble is so low, on the order of $10^{-3}$
atoms cm$^{-3}$, that only
about $3~10^{-5}$ \Msun\ has been swept up. This is negligible
compared to the mass of the nebula (Smith et al. 1997).


On the other hand,
the stellar wind that is blowing since the nebula was ejected
may have caught up with the nebula and may have changed the 
composition of the nebula from the inside. 
The amount of wind material that has run into the nebula is
\begin{equation}
\label{eq:windmass}
\Delta M \simeq \Mw \left( \tau_{\rm dyn} - \frac{r_{\rm in}}{v_{\rm w}}\right)
         \simeq \Mw \tau_{\rm dyn} \left( 1 - \frac{v_{\rm exp}}{2 v_{\rm w}}
         \right)
\end{equation}
where \Mw\ and $v_{\rm w}$ are the mass loss rate and
the velocity of the wind since the ejection of the
nebula; $v_{\rm exp}$ and $\tau_{\rm dyn}$ are the 
expansion velocity and the  dynamical age of the nebula and 
$r_{\rm in}\simeq 0.5~ r_{\rm neb}$ is the inner 
radius of the nebula. 
The mean mass loss rate \Mw\ is not very well known for LBVs
because the mass loss rate varies strongly between phases of 
high and low mass loss rates.
Studies of mass loss in various LBV phases indicate that the
mean mass loss rate of AG Car is  $\Mw \simeq 8~10^{-5}$ \Msunyr\
and that the mean wind velocity is $v_{\rm w} \simeq 110$ \kms\
(Leitherer 1997). With a nebular expansion velocity of 70 \kms\
and a dynamical age of $1~10^4$ years (Table \ref{tbl:parameters})
we find that $\Delta M = 0.5 \Msun$. A similar estimate for
R\,127 with $\Mw \simeq 6~10^{-5}$ \Msunyr, $v_{\rm w}=110$ \kms,
$v_{\rm exp}=28$ \kms\ and $\tau_{\rm dyn}=2~10^4$ years
gives $\Delta M = 1.0$ \Msun. These values are upper limits
if the stars have not been LBVs immediately 
since the ejection of the nebula.
For instance, if the nebula
was ejected during the RSG phase and the star evolved to
an LBV since then, the average wind velocity (and possibly also the
average mass loss rate) will have been lower. 

Garc\'ia-Segura et al. (1996b) have calculated the evolution
of a nebula that was ejected during the RSG phase by a star with
$M_i=35 \Msun$ and interacted with the wind produced in the subsequent
Wolf-Rayet
(WR) phase. In their model the evolution from RSG to WR star is very 
short and lasts only about 100 years. This evolution does not 
allow the presence of an LBV phase about $10^4$ years after
the nebula ejection, so it is not applicable to our program stars.  

We conclude that the initially ejected nebula may have mixed
with wind material after the ejection. 
If this material
had the same composition as the nebula, no enrichment has occurred.
On the other hand, if the wind material had a
composition different from that of the ejected nebula, the 
presently observed chemical abundance of the nebula is a mixture
of the two components. 
We will later present  evidence (Sect. 7) that 
an enrichment of the surface layers may indeed have occured after the
nebula was ejected. This may have resulted in a slight enrichment of
the nebula after its ejection. 


\section{The expected nebular abundances due to CNO-enrichment}
\label{sec:4}

The abundance ratios of N/O and N/S, 
derived from observations of LBV nebulae and
given in Table \ref{tbl:parameters}, can be used to derive the evolutionary
stage of the star when the nebula was ejected.
 The N/O ratio in LBV nebulae is about 1 to 6
whereas we only have an upper limit for the N/S ratio  of $<$ 30 to
80, except for the P Cygni nebula (Table \ref{tbl:parameters}).
These ratios will be compared with predictions from evolutionary
calculations for stars with convective and rotation induced mixing.
Although we have no He/H ratio determination for the nebulae, we will
also study  the changes in the He/H ratio in the stars during their
evolution. This is of interest because the data in Table \ref{tbl:parameters}
show that the {\it photospheres}
of LBVs are remarkably helium-rich, and have a narrow
number ratio of 
He/H $= 0.4$ to 0.7 or a helium mass fraction of $Y \simeq 0.62$ to 0.73.
 
Evolutionary models of massive stars and the resulting variations
in the surface abundances have been calculated by several groups,
e.g. Maeder \& Meynet (1989), Langer (1991), Schaller \etal\ (1992).
We will use the evolutionary calculations by Meynet et al. (1994)
and Schaerer et al. (1996).
The evolution is strongly affected by mass loss and since different
authors used different expressions for the mass loss rates, their
evolutionary tracks can differ greatly. For instance, whether a massive star
will go through a red supergiant phase or not
depends on the amount of mass
lost during the main sequence and the blue supergiant phase.

In the next sections we describe the expected abundance
changes in the interiors of massive stars from a very basic point of view.
The purpose of this description is to point out where the layers with
the observed abundance ratio N/O $\simeq 1$ to 6
are located in the star,
and at what point in the evolution they may have been ejected
to form the LBV nebulae. We also determine the location of the
layers where He/H $\simeq 0.4$ because they appear
at the surface of the star {\it after} the nebula was ejected.
We will concentrate our efforts on
stars with initial masses in the range of 85 to 40 \Msun, i.e.
roughly the range of the LBVs.
However, since rotating 20 \Msun\ models potentially
can evolve to luminosities of more than $3 \times 10^5$ \Lsun\
(Heger et al. 1997),
we will also study the evolution of a star with an initial mass of 20 \Msun.
For the prediction  of the changing surface abundances 
we need to know the chemical evolution inside the star
and the size of the convective core. For this we adopted the 
evolutionary calculations of the Geneva group (Meynet et al. 1994,
Schaerer et al. 1996). These authors calculated evolutionary tracks 
for two different sets of mass loss rates on the main sequence:
the ``normal'' rates (de Jager et al. 1988) and the same rates 
enhanced by a factor 2. Meynet et al. (1994) 
argued that the tracks with the 
higher mass loss rates agree better with the observations.
Therefore we adopt the internal structure of the evolution models  
of Meynet et al. (1994) with the enhanced mass loss rates. 

The initial abundances in terms of mass fractions in the Z=0.02 (Galactic)
models of Meynet et al. (1994) for C, N and O are
$4.86\times 10^{-3}, 1.24 \times 10^{-3}, 1.05 \times 10^{-2}$ which are essentially solar.
This corresponds to number ratios of

\begin{eqnarray}
\label{eq:abundances}
 N/C/O ~ &=& ~ 1.0/4.6 / 7.5
\end{eqnarray}
For lower metallicity models of Z=0.008 Meynet et al. (1994) scaled
the absolute abundances, but kept the same relative ratios.

It is well known that H\,II region abundances are lower than solar values.
The average H\,II region abundances of Shaver et al. (1983) give the following
number ratios:

\begin{eqnarray}
\label{eq:abundances-HIIG}
 N/C/O/S~ &=& ~ 1.0/7.8 /13/0.31 ~~~~~~~~~~{\rm for~Galactic~nebulae}
\end{eqnarray}
The nebular N/O ratio is a factor 0.52 smaller than 
assumed in the evolutionary models. 
This difference is even worse for the LMC (Z=0.008) since
the H\,II region analyses of Dufour (1984) and
Russell \& Dopita (1990) give number ratios of 

\begin{eqnarray}
\label{eq:abundances-HIILMC}
 N/C/O/S~ &=& ~ 1.0/7.6 /24/0.59 ~~~~~~~~~~{\rm for~LMC~nebulae}
\end{eqnarray}
So the N/O ratio of the H\,II regions in the LMC is a factor
0.31 smaller than adopted in the evolutionary calculations for
the LMC stars with $Z=0.008$. 
Although the question of the correct initial abundances
is still controversial since B-stars sometimes give different results to the
H\,II region analyses (Killian 1992, Rolleston et al. 1996, Venn 1999)
we believe it is more
correct to adopt the H\,II region abundances as the initial
abundances of the stars. Therefore we adopt the relative initial
abundances of Eq. (\ref{eq:abundances-HIIG}) for Galactic stars
and Eq. (\ref{eq:abundances-HIILMC}) for the LMC stars. We 
re-normalised the initial stellar abundances of the evolutionary
models of Meynet et al. (1994) that we will discuss in the following
sections. Obviously, this scaling only applies to the {\it initial}
abundances; the abundances of the CN or ON-cycle products are not
sensitive to the initial abundance ratios, except when these cycles
have not yet reached equilibrium.
 
The abundance ratios of the gas enriched by the
CNO-cycle in equilibrium in a 60 \Msun\ star are
 
\begin{eqnarray}
\label{eq:CNO-equil}
 N/C/O ~ &\simeq& ~ 1.0 /0.035 / 0.017 ~~~~~
{\rm for~CNO-equilibrium~abundances}
\end{eqnarray}
These ratios are not affected by the change in the adopted
initial composition described above.  
Using the abundance ratios of Eq. (\ref{eq:CNO-equil}), it is easy to
show that 
for a mixture consisting of a mass fraction $1-f$ of initial
composition and a mass fraction $f$ of the CNO-equilibrium
material we find that $f \simeq 0.35$ to 0.8 for 
the observed ratio of 1$<$ N/O $<$5 in the LBV nebulae.
This admittedly rough estimate shows that a significant fraction
of the gas in the LBV nebulae must have gone through CN or
CNO processing.  

 We conclude that the abundances in LBV nebulae can only be explained if the
ejected material was a mixture of original gas and CN
or CNO-processed gas.
This conclusion was already reached by Smith et al. (1998), based
on the N/H and O/H ratios.
The mixing must have occurred in the star {\it before} the nebula was ejected.
In the next two sections we will consider two mechanisms for the
mixing: convective mixing and rotation induced mixing.


\section{The surface abundances during the red supergiant phase}
\label{sec:5}

In this section we discuss the effect of convective mixing 
in the stellar core during the main sequence phase and of
convective envelope mixing during the red supergiant phase.
We calculate the expected surface abundance of H,
He, C, N and O in the red supergiant phase for different
mass loss rates in the main sequence phase.

\subsection{The global picture}
\label{sec:5a}

The internal structure of a massive star during the main sequence phase
and at the start of the core He-burning is sketched in the
Kippenhahn-diagram of Figure 1. The left panel 
shows the internal structure as a function of time (horizontal axis)
and mass fraction compared to the initial mass (vertical axis). 
During the main sequence it consists  
of the following layers, from the inside to  the outside: \\
(a) a core where H is converted into He by the CN-cycle and the ON-cycle. \\
(b) a convective core that extends beyond the H-burning zone. This 
includes the overshooting region. The mass of the convective core 
decreases with time.\\
(c) possibly a thin layer where the overshooting has been less effective
and where the chemical mixing with the lower layers is incomplete.
(It is unlikely that the ejected LBV nebula comes from this thin
region of possibly incomplete overshooting. Firstly, this layer will be thin
as it is expected to extend over  a small fraction of a pressure
scale height during the early main sequence phase. Secondly,
the ejection of exactly this layer in all LBV nebulae would require a strict
fine tuning of the mass loss rate, because the star must have lost
just enough mass during the previous phases to get rid of the
original layer, but leave the mixed layer unaffected until it is 
blown out to form the nebula.)\\
(d) the envelope in radiative equilibrium. 
The upper part of the envelope 
has the original composition. The lower layers of 
the envelope have a variable composition due to mixing by 
the receding convective core. Mass loss removes part of the top layer.\\

The CN-cycle in the core reaches equilibrium very quickly, within a 
few $10^5$ years, and so the initial N/C ratio of 0.13 will be
increased to about 30 throughout the layers of the initial convective core.
The ON-cycle reaches equilibrium more slowly,
typically in a few $10^6$ years, so the O-abundance slowly decreases
during the main sequence phase and the N/O ratio slowly increases.
Since the convective core is shrinking in mass, the N/O ratio 
at the end of the main sequence phase will increase with depth
in the layers between the extent of the initial 
and the late convective core.

The right panel  of 
Figure 1 schematically shows the
chemical composition of N throughout the star
before the envelope convection sets in (left)
and during the red supergiant phase when the convection has set in (right).
The amount of mass of the original layer at the time 
when the outer convection starts
depends critically on the mass loss during the earlier phases. 
During the RSG phase N is destroyed in the core by the He-burning.
Above it is a region with CN-equilibrium material. 
The outer convection layers consist of mixed gas. The composition of this
layer will critically depend on the ratio between the mass of the  
original layer and the mass of the CN-layer that is reached by the
outer convection.
We will calculate this composition below.

\begin{figure}
\centerline{\psfig{figure=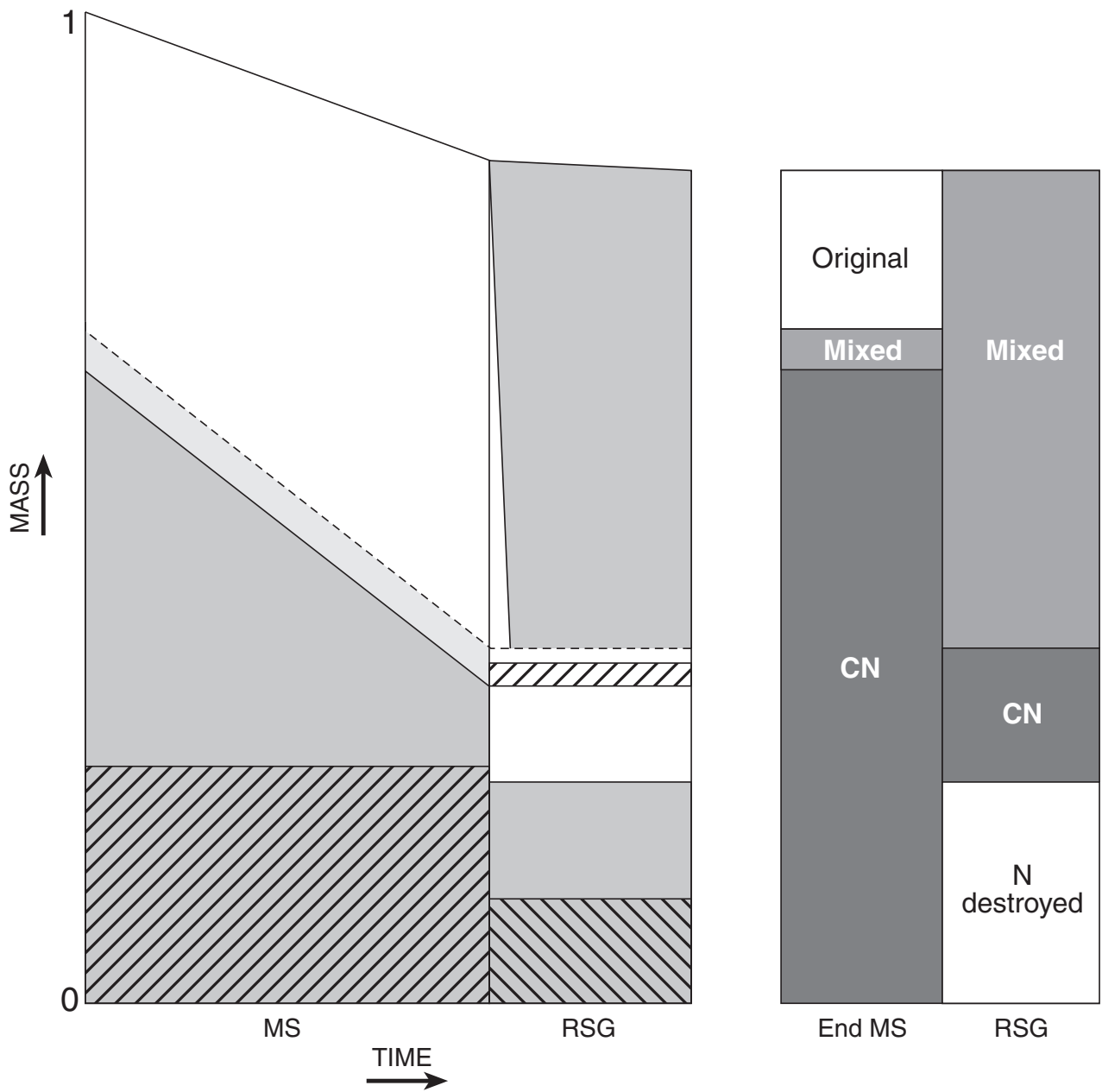,width=\columnwidth}}
\caption[]{Left panel: a schematic diagram of the internal structure
of a massive star during the main sequence phase and the early
red supergiant phase. The vertical scale is the mass fraction,
relative to the initial mass. Grey regions are convective.
Slanted dashed regions are the H-burning (forward slanted) and the
RSG He-burning zone (backward slanted). The layer on top of the convection zone
may have additional incomplete mixing (light-grey). The time scale of the 
He-burning phase is exagerated.\\
Right panel: the chemical composition of N at the end of the
main sequence and after the onset of the convection in the outer
envelope. Dark grey areas have CN-equilibrium composition, 
light grey areas have a mixed composition (adapted from
Maeder \& Meynet, 1987)}
\label{fig:Kippenhahn}
\end{figure}

\subsection{The internal abundance prior to the red supergiant phase}
\label{sec:5b}

The abundance pattern of C, N and O in the star just prior to the
red supergiant phase is shown in Figure 2
for models with Z=0.02 (Galactic) and
Z=0.008 (LMC) for stars of initial masses of 85, 60, 40 and 20 \Msun.
The abundance structure was derived from the
evolutionary models of Meynet et al. (1994) for enhanced mass loss
rates,
but corrected for the initial CNO abundances from the H\,II regions.
The figure shows the enhancements or depletions of 
CNO in the star as a function of the fraction of the {\it initial} mass. 
 The models show that the N abundance reaches it CN-equilibrium
value nearly throughout the whole region of the initial convective core.
The C abundance shows a steep drop at the edge of the initial convective
core, but then slowly rises by about a factor 2 to the inside. 
This is because the ON cycle slowly decreases the O abundance and increases the
C abundance as the convective core shrinks in mass.  
In comparing the abundance distributions of the two metallicities,
we see that the abundance patterns are very similar. 
The main difference is in the abundances of O, which has a steeper profile
in low metallicity stars, because the 
ON-cycle more quickly reaches equilibrium in the stars with Z=0.008 than with
Z=0.02. For stars of lower metallicity the C, N and O nuclei have to 
go through the CNO nuclear cycle more frequently than for higher metallicity
in order to reach the same conversion rate of H into He. 
We will use these chemical profiles to calculate the composition
of the envelope after the convective mixing has set in.

\begin{figure}
\label{fig:modelabundance}
\centerline{\psfig{figure=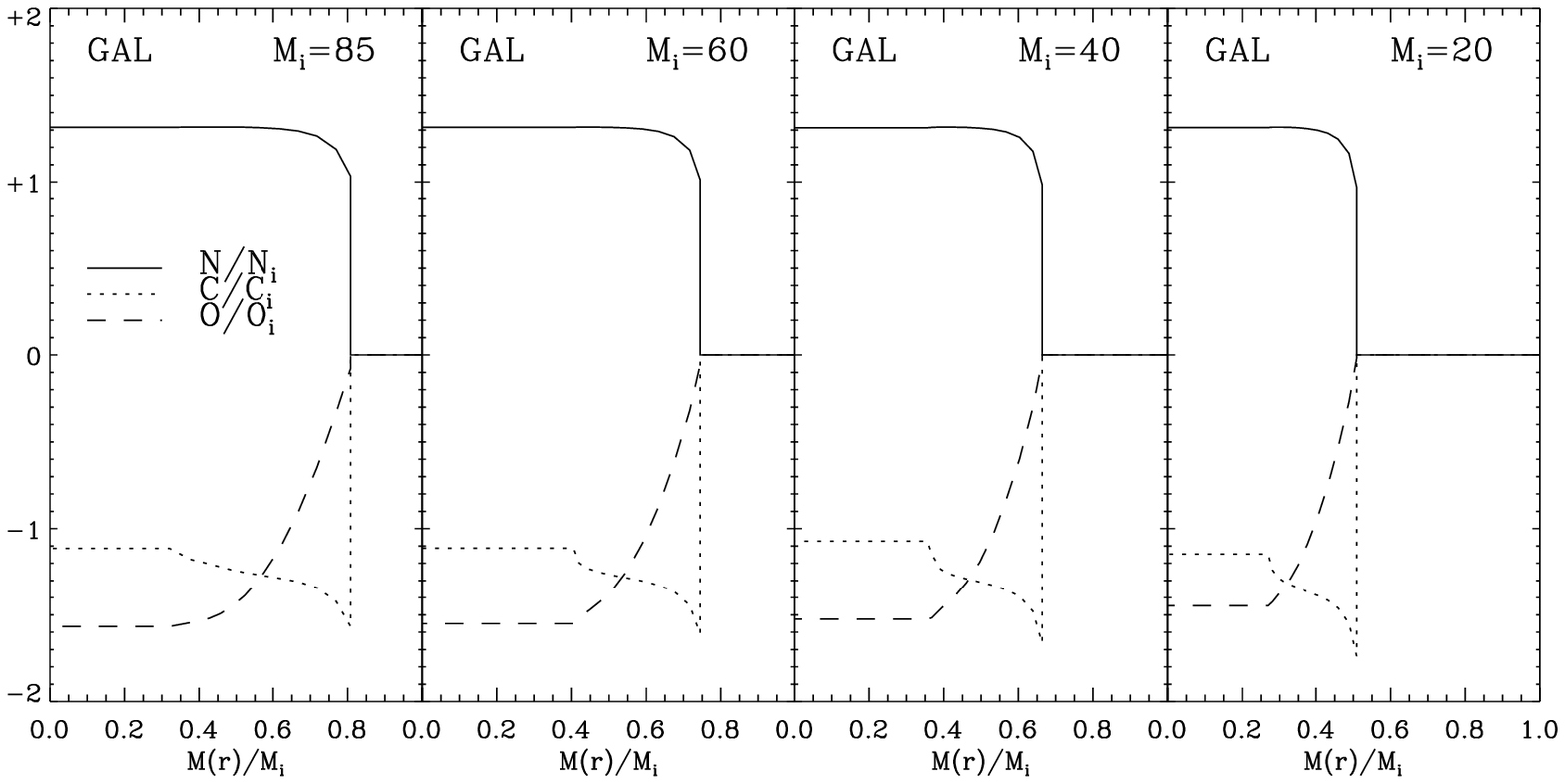,width=\columnwidth}}
\vspace{-3.5cm}
\centerline{\psfig{figure=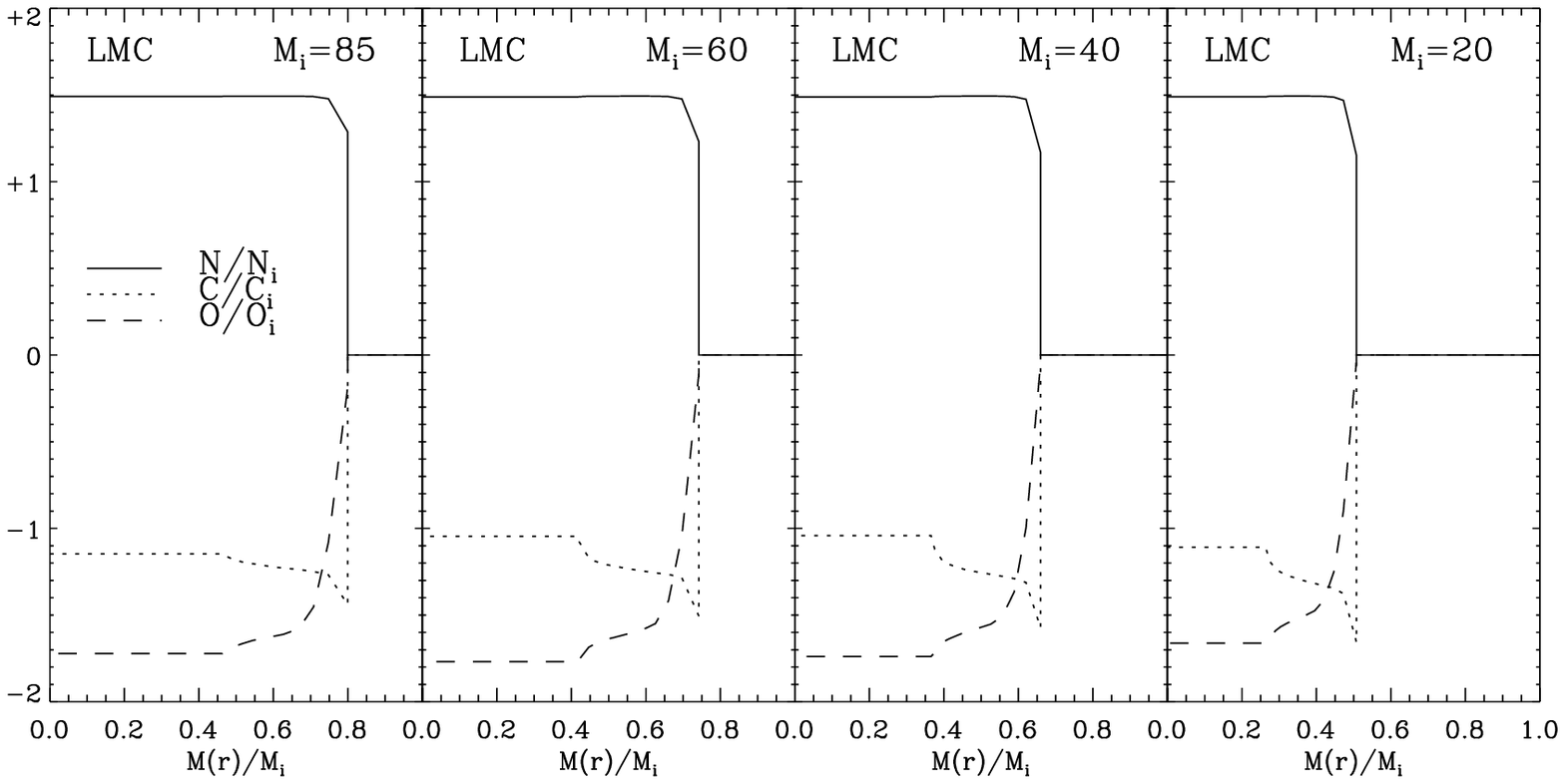,width=\columnwidth}}
\vspace{-1.0cm}
\caption[]{The predicted abundance pattern at the end of the core H-burning
phase as a function of the fraction of the initial mass, for models with
inital masses of 85, 60, 40  and 20 \Msun, with metallicities of
Z=0.02 (top panel) and Z=0.008 (lower panel). The vertical axis shows
the logarithmic over- or underabundance, compared to the initial
value. The data are  derived from the evolutionary
models of Meynet et al. (1994) for enhanced mass loss rates, modified
for the initial abundances of H\,II region composition. }
\end{figure}

We remind the reader that the {\it internal} chemical evolution of the stars during the
pre-RSG phase is very little affected by the mass loss. 
The main effect of the 
mass loss on the internal evolution is through the luminosity: a star with
a higher mass loss rate will more slowly increase its luminosity during the
main sequence phase than a star with a low
mass loss rate, so the main sequence phase of the star with the higher
mass loss rate will last longer.


\subsection{The envelope abundance after convective mixing in the RSG phase}
\label{sec:5c}


A considerable fraction of the stellar mass may have been lost
before the star reaches the onset of the convective mixing. 
This depends on the mass loss rate during the earlier phases
and hence on the metallicity. 
Let us assume that a certain fraction 
$\qmdot=\Delta M / M_i$ of the initial mass, $M_i$, was lost
before the onset of the convection. 
Because the mass loss 
prior to the red supergiant phase is not well known (in the
evolutionary calculations a very high mass loss rate in the blue
supergiant phase is {\it assumed} almost arbitrarily to prevent the
most massive stars from evolving over to the RSG phase!), 
we will treat \qmdot\ as a free parameter. 
Let us express  
the amount of mass in the core that is not reached by  the envelope convection
as $q_{\rm core}~ M_i$. We set this value to the
mass of the convective core 
at the end of the core H-burning phase, plus the few solar masses
of the shell H-burning. From the evolutionary models we find that
this shell-burning mass is approximately 0.05 of the initial mass. 
We adopt this value.

Having specified the chemical profile (Fig. 2),
the mass of the star at the onset of the envelope convection,
$(1-\qmdot)M_i$, and the mass of the core that does not mix,
$q_{\rm core}M_i$, we can calculate the abundance of the mixed envelope.
Figure 3 shows the predicted
abundance pattern of the convective envelope as a function of
the remaining mass of the star at the onset of the red supergiant phase,
i.e. $M_{RSG}/M_i= \{1 - \qmdot\} $.
For $M_{RSG}/M_i =1$, that is for no mass loss (\qmdot=0), 
the mass of the layer with the 
original composition has its maximum value. 
This gives the minimum 
abundance changes in the red supergiant phase. When \qmdot\
is so large that there is no layer of original composition left at the
onset of the envelope convection, the abundances converge to the
CN equilibrium values, except for the fact that the NO cycle may not have
reached equilibrium. 
Notice the very steep dependence of the N/O ratio on the amount of mass
that is lost during the main sequence phase, especially in the region
where N/O$\simeq 3$ to 30.

Figure 3 shows that
 the N/O ratio rises more steeply for the Z=0.008 models
than for the Z=0.02 models. This is due to the steeper initial decrease of the
O-abundance during the early MS-phase when the convective
core  has its maximum extent. Therefore the O-abundance is lower in the 
Z=0.008 models than
in the Z=0.02 models over a larger fraction of the stellar mass
(see Figure 2). 
For a given initial mass the N/O ratio rapidly increases with 
decreasing mass of the star at the onset of the
RSG phase, i.e. with increasing mass loss.
For Z=0.02 the mass loss rates adopted in the evolutionary 
calculations of
Meynet et al. (1994) leave 38 \% of the initial mass for $M_i=85 \Msun$,
65 \% for 60 \Msun\ and  80 \% for 40 \Msun. 
From the figure we read
that this would predict an N/O ratio during the RSG phase
of log N/O=1.75, 1.35 and 0.35 for 85, 60 and 40 \Msun\ respectively.
If the mass loss rate prior to the RSG phase were higher, then the
ratios would by higher.
For Z=0.008 the evolutionary models predict a mass at the end of the
main sequence of 68 \% for $M_i=85$, 76 \% for $M_i=60$ and 89 \%
for $M_i=40~ \Msun$. The correspondingly predicted N/O ratio after convective
envelope mixing is log N/O = 1.75, 1.00 and 0.20 respectively.

\begin{figure}
\centerline{\psfig{figure=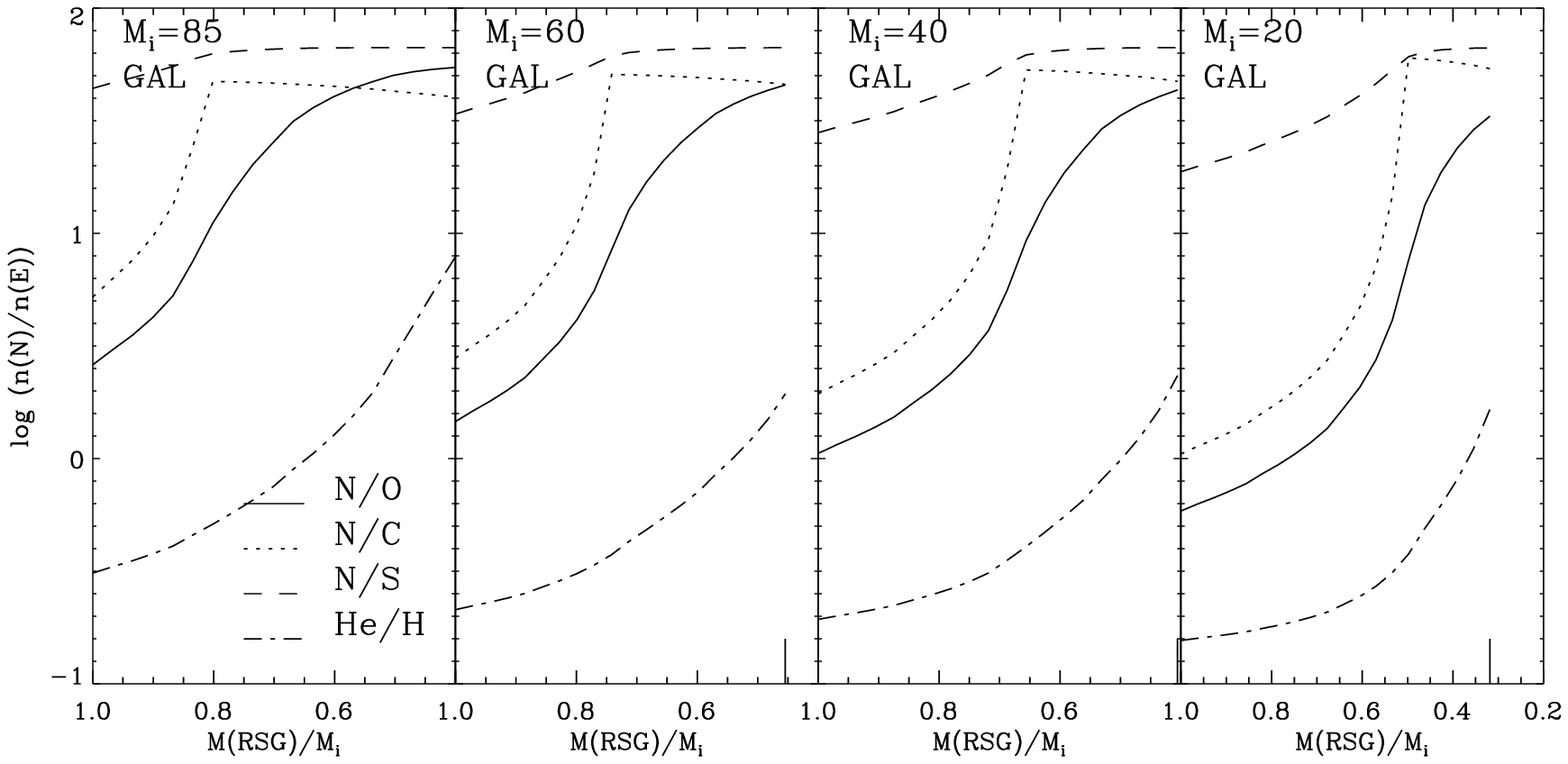,width=\columnwidth}}
\vspace{-3.5cm}
\centerline{\psfig{figure=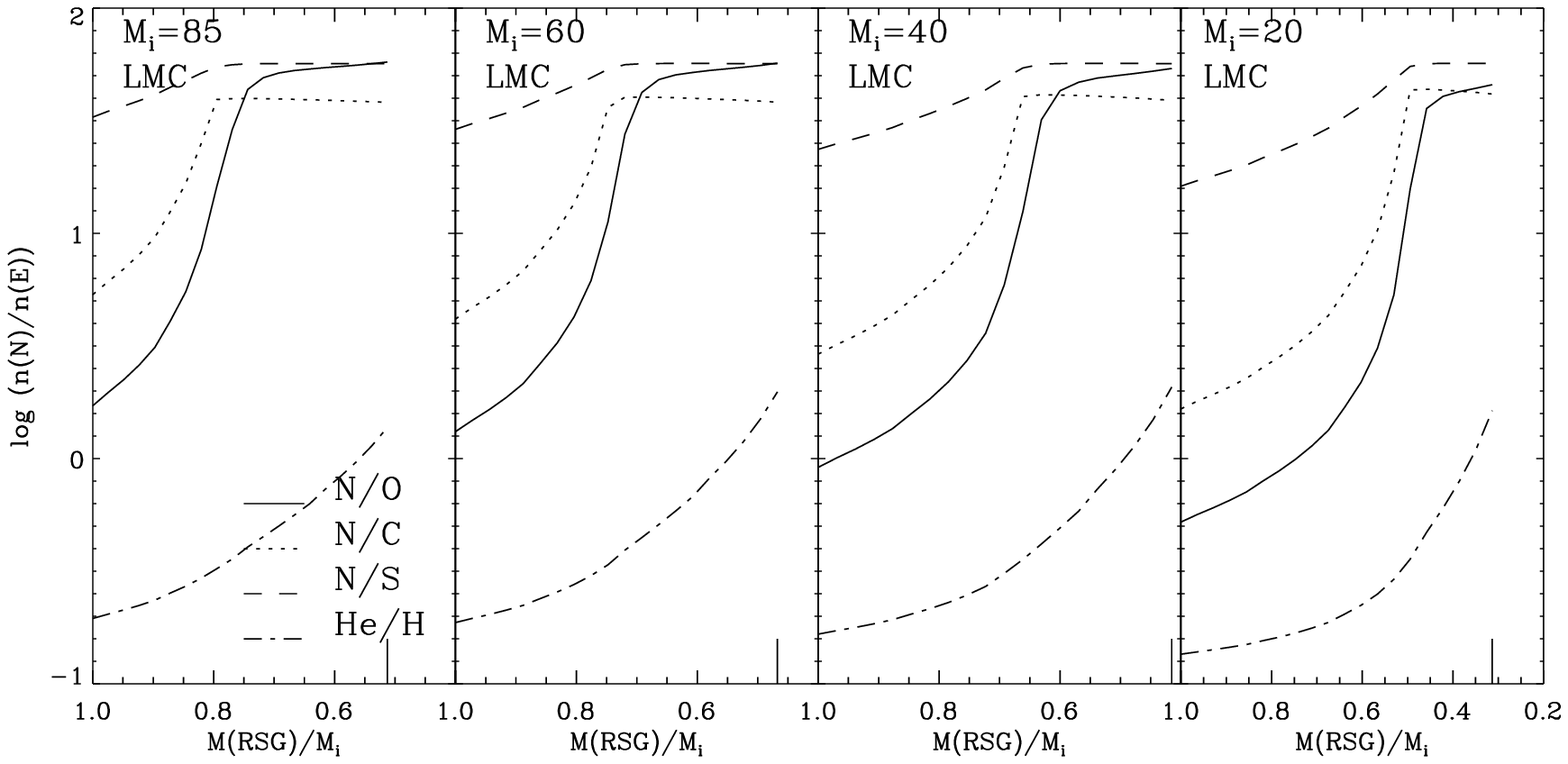,width=\columnwidth}}
\vspace{-1.0cm}
\caption[]{The logarithmic number ratios of N/O, N/C, 
N/S and He/H as a function 
of the remaining fraction of the mass of the star at the onset of the
outer convection. 
The long vertical tickmark gives the core mass fraction
that does not partake in
the convection. The upper figure is for Z=0.02 and the lower figure
is for Z=0.008. (From the evolutionary models of Meynet et al. (1994)
but with the relative initial CNO abundances derived from the H\,II
regions; see text).}
\label{fig:rsgmix}
\end{figure}

We want to point out that the surface abundances 
predicted in the Geneva models (e.g. Meynet et al. 1994) during
the RSG phase are closer to the initial abundance than in our calculations.
This is due to the fact that the convection is artificially reduced
in the Geneva models of the red supergiants, in order to avoid a  large
density inversion that causes numerical problems. This reduction
of the convection produces in the Geneva tracks red supergiant models 
which have too high
effective temperatures: the convection in the models does not reach deep
enough, so the convective mixing is reduced, the envelope 
is not extended enough and the radius is not large 
enough to make cool red supergiants (Maeder, Private Communications).

\subsection{Comparison with the observations}
\label{sec:5d}

We compare the observed nebular abundances with those expected from 
stars in the RSG phase as shown in Figure 3. 
If we assume that the nebula was ejected during the 
RSG phase, we can derive the total amount of mass that was lost from 
the star before the onset of the outer convection. The results
are summarized in Table 2. 
The stars and their initial mass are listed in Columns 1 and 2. Column
3 gives the observed abundance ratio that was used to
derive the mass of the star when it entered the RSG phase, $M_{RSG}$.
This mass is compared to the initial mass in Column 4. Column 5 gives
the mass of the star that was lost prior to the RSG phase, derived
from the observed abundance ratio and column 6 gives the mass lost on the main
sequence adopted in the evolutionary calculations by Meynet et
al. (1994). Column 7 gives the nebular He/H ratio that is predicted
on the basis of the observed N/O or N/S ratio, using the results
in Figure 3. The last column gives the observed photospheric
He/H ratio.

\begin{itemize}

\item {\bf AG Car}:
The luminosity
of this star suggests an initial mass of about 60 \Msun\ if the star
is presently in the post-RSG phase. 
  From Fig. 
3 we see that the observed ratio N/O $=6 \pm 2$ 
can be explained by
convective mixing if $M_{\rm RSG}/M_i=0.78 \pm 0.03$ i.e. the star has lost 
22$\pm 3$ percent of its mass, corresponding to 13.0 $\pm$1.8 \Msun,
prior to the onset of the convection.
The evolutionary calculations by Meynet et al. (1994) for enhanced mass loss 
predict that the star
has lost about 35 percent of its mass at the end of the 
H core burning phase, and 20 percent for normal mass loss (see
Sect. \ref{sec:4}).
The empirically derived value is close to the predicted value for
normal mass loss rates.
We see that the nebular N/O ratio can be explained if the 
nebula was ejected during the RSG phase and the star had lost
11 to 15 \Msun\  prior to this phase.
The surface He/H ratio for a star of $M_i = 60 \Msun$ 
that has lost 13 \Msun\ before the onset
of the RSG phase is He/H$=0.32 \pm 0.03$. The observed surface ratio of
0.4 for the star AG Car is slightly higher than this. 
This  might indicate that the star has lost more mass
since the nebula was ejected, bringing more enriched layers
to the surface in the blue supergiant phase.

\item{\bf P Cygni}:
If P Cygni is in the post-RSG phase its initial mass must have been
about 50 \Msun. Therefore we interpolated the predicted N/S and the
N/H ratio for the two models of 40 and 60 \Msun. Unfortunately we only
have an observed lower limit to the N/O ratio (Table 1). This
lower limit does not set a useful constraint on the mass of the RSG.
Therefore we use the N/S ratio, which is however a less senstive 
indicator of the fraction of mass that is lost than the N/O ratio.
Moreover, the derived N/S ratio is uncertain (see Sect.  2).   
The observed ratio of N/S$\simeq 33$ corresponds to
$M_{RSG}/M_i=1.0$ for $M_i=60$ \Msun, and $M_{RSG}/M_i=0.87$ 
for $M_i=40$ \Msun.
This suggests that {\it if} the nebula was ejected during the RSG phase,
the star of initially 50 \Msun\ must have lost about 7
percent of its mass prior to this phase. The observed N/S ratio 
corresponds to a predicted He/H ratio of 0.22, which is much
smaller than the photospheric ratio of He/H=0.42.
The evolution calculations with enhanced mass loss
of Meynet et al. (1994) show 
that a star of 50 \Msun\ is expected to
lose about 
30 percent of its mass during the core H-burning. 
This is significantly higher than the (uncertain) value of 7 \%
that we derived from the N/S ratio. 
For a mass loss of 30 percent we predict an N/S ratio of 50, which may
still be within the uncertainty of the derived abundances.

\item{\bf R\,127}:
From the location of this star in the HR-diagram we find that
the initial mass must have been about 75 \Msun\ 
if the star is now in the post-RSG phase.
Comparison with the results in Fig 3  for the 
Z=0.008 models shows that the observed abundance ratio 
of N/O = 0.9 $\pm 0.4$ 
is smaller than the predicted one, even if the star had not lost
any mass at all prior to the onset of the envelope convection!
The predicted He/H ratio for $M_{RSG}/M_i=1.0$ is 0.2, which is
much smaller than the present photospheric ratio of 0.50. 
The evolutionary tracks with enhanced mass loss of Meynet et al.
 (1994) predict that the star has lost 25 percent of its mass
during the core H-burning and that the N/O ratio should be larger
than 100, which is much higher than observed. 
Assuming that the mass loss rate of R\,127 has been about 0.4
times 
that of its Galactic counterpart AG Car, because the mass loss rate 
is proportional to metallicity as $Z^{0.86}$ (Vink et al. 2000), 
we would expect a
$M_{RSG} \simeq 0.91~M_i = 68~\Msun$, which corresponds
to N/O $\simeq 3 $. This is twice as high as the
observed maximum ratio of 1.3. This result depends sensitively on the
adopted initial mass of the star. If the luminosity was 6.0 dex
instead of 6.1
(which is within the accuracy of the luminosity
determination), 
the initial mass would be 60 \Msun\ and the predicted
N/O ratio would be 1.5, which is only marginally higher than the
upper limit of the observed ratio.

\item{\bf S\,119}:
The initial mass, estimated from the stellar parameters, is 
$M_i \simeq 60~ \Msun$ if the star in presently in a post-RSG phase.
Comparison of the observed nebular ratio of N/O $= 1.9 \pm 0.5$
with the results in Fig \ref{fig:rsgmix} for 60 \Msun\
shows that the star must have lost 7 percent of its mass if N/O=1.9
and 12 percent if N/O=2.4 before the envelope convection started.
The lower limit of the observed ratio
of N/O=1.4 is not compatible with the predictions.
The predicted He/H ratio for 7 to 12 \% mass loss is about 0.20. This is
much smaller than the present photospheric He/H ratio of 0.67. 
The evolutionary models of an LMC 60 \Msun\ star of $Z=0.008$  
by Meynet et al. (1994) show that the star has lost about 25
percent of its mass prior to the RSG phase. 
However, if that were the case the N/O ratio of the nebula
should have been about 10, which is clearly incompatible
with the observed value.

\end{itemize}

We conclude that 
the N/O abundance ratio in the nebula is a very sensitive indicator of the 
mass that was lost prior to the onset of the envelope convection, {\it
  if} the 
nebula was ejected during the RSG phase. 
The observed ratio of AG Car is in agreement with
the evolutionary predictions for the mass that the star has 
lost at the end of the main sequence phase if the ``normal'' mass loss
rates of de Jager et al. (1988) are adopted.
For the P Cygni nebula
we used the observed uncertain N/S ratio because for the N/O ratio only
a lower limit is known. The uncertain N/S ratio is about 40 percent
smaller than the value predicted by stellar evolution. 
For the LMC stars the N/O ratios point to a mass loss rate
that is smaller than adopted in the evolutionary tracks by Meynet et
al. (1994) if the nebulae were indeed ejected during the RSG phase.  

For all stars we find that the predicted He/H ratio, that is
associated with the observed N/O or N/S ratio is smaller than the
observed photospheric He/H ratio. The difference is small for AG Car,
but it is significant for P Cygni and R\,127.  Taken at face value, this shows that a significant
enrichment of the surface layers must have occured {\it after} the
LBV nebula was ejected. We return to this in Sect. \ref{sec:7}.
 
We remind the reader that in these calculations 
we have adopted a considerably deeper mixing  
during the RSG phase than in the evolutionary calculations of the
Geneva group. In the Geneva models the deep convection is trimmed to avoid
numerical instabilities in the code, which results in less severe
mixing than in our simplified models.


\section{The surface abundances of stars with rotation induced mixing}
\label{sec:6}

In this section we consider the effects of rotation induced mixing
on the chemical evolution of the star. Evidence for envelope
mixing during the main sequence phase comes from the
existence of OB supergiants with enhanced N abundance
(Walborn, 1988),
 and also from the boron depletion in their photospheres
(cf. Fliegner et al. 1996).
This indicates that in these stars the products of the
nuclear fusion already have reached the stellar surface
at or near the end of the main sequence phase.
This is not surprising. Firstly, the mean rotation velocity of the
main sequence O-stars is high and of order several hundred \kms.
Secondly, the grid of stellar evolution models for stars in the mass range
10 to 25 \Msun\ of Heger et al. (2000) shows that the rotational mixing
is more efficient for higher initial masses, as radiation pressure
becomes more dominant and renders the influence of mean molecular
weight barriers less important (see also Maeder 1998).

Therefore we study the effect of mixing on the surface abundance
at the end of the main sequence phase.
We will calculate the surface abundances as a function of the
ratio between the mixing timescale \tmix, and the timescale for
core H-burning, \taums. Before presenting the calculations
we discuss the mixing timescale that might be expected for rotation.
 
\subsection{The time scale for rotation induced mixing}
\label{sec:6a}
 
Rapidly rotating stars will develop an Eddington-Sweet circulation
due to thermal imbalance (e.g. Kippenhahn \& Weigert, 1990, p 439). 
This circulation will bring nuclear products
from the convective core into the non-convective envelope, where they may
appear at the surface of the star. In differentially rotating stars,
among other processes, the
shear instability
can produce turbulence which can also result in chemical mixing
(Maeder 1998).
Assuming that mean molecular weight barriers have only a limited
efficiency in preventing the mixing
(Maeder \& Zahn 1998, Heger, Langer \& Woosley 2000),
the timescale for the Eddington-Sweet circulation, which 
dominates the mixing time scale for massive main sequence stars,
is related to the Kelvin Helmholtz timescale (Kippenhahn \& Weigert,
1990, p 439),

\begin{equation}
\label{eq:circulation}
\tau_{\rm circ}~=~\tau_{\rm KH} / \chi
\end{equation}
where $\tauKH=G M_*^2 /R_* L_*$ 
is typically about $10^{-2}$ times the main sequence lifetime,
and
 
\begin{equation}
\label{eq:chi}
 \chi~=~ \frac{2}{3}\frac{g_{\rm centr}}{g_N} =
\frac{2}{3} \frac{\omega^2 R_*^3}{G M_*}
\end{equation}
where $g_{\rm centr}$ and $g_N$ are the centrifugal acceleration and the
acceleration of gravity at the equator and $\omega$ is the
angular velocity (Kippenhahn \& Weigert, 1990). These two expressions are 
for
a solidly rotating star, which is a good approximation for the considered
case.  
While recent, more sophisticated models of rotating massive main
sequence stars show that the rotational mixing is really determined
by an interplay of Eddington-Sweet currents, shear mixing, and
baroclinic turbulence (Heger et al. 2000, Meynet \& Maeder 2000),
they confirm the basic trends predicted by Eq.~(6).
We see that $0 < \chi < 1$ and that the circulation time
is on the order of the Kelvin-Helmholtz time for
stars near critical rotation speed and increases rapidly for
slower rotation.
 
The main sequence lifetime, the mass and
radius during the ZAMS phase of stars with $40 < M_i < 85~ \Msun$,
derived from the evolutionary calculations for Z=0.02 by
Meynet et al. (1994) depend on the
initial mass approximately as
 
\begin{eqnarray}
\label{eq:taums}
\taums       &\simeq& 3.1~10^{7}~M_i^{-0.544} \nonumber \\
L_*          &\simeq& 2.1~10^{2}~M_i^{+1.917} \nonumber\\
R_*          &\simeq& 0.93~M_i^{+0.599}
\end{eqnarray}
where \taums\ is in years and all stellar quantities are in solar units.
For $Z=0.008$ the luminosity is about 10 \% lower,
the main sequence lifetime is about 10 \% larger,
 and the radius is about 10 \% smaller.
Combining
Eqs. (\ref{eq:circulation}), (\ref{eq:chi}) and  (\ref{eq:taums})
we find an expression for the
ratio between the circulation time and the MS lifetime
for stars with $40 <M_i < 85~ \Msun$ of
 
\begin{equation}
\label{eq:mixingtimegal}
\frac{\tau_{\rm circ}}{\taums} \simeq
\left(\frac{P}{5.68}\right)^2 \left( \frac{M_i}{60}\right)^{-0.77}
\end{equation}
for $Z=0.02$, where the rotational period $P$ is in days.
For $Z=0.008$ the ratio is about
 
\begin{equation}
\label{eq:mixingtimelmc}
\frac{\tau_{\rm circ}}{\taums} \simeq
\left(\frac{P}{4.82}\right)^2 \left( \frac{M_i}{60}\right)^{-0.67}
\end{equation}
The dependence of $\tau_{\rm circ}/\taums$ on $M_i$ is slightly
flatter for $Z=0.008$ than for $Z=0.02$.
In deriving these expressions we have used the luminosity and the
radius of the ZAMS and we have adopted solid rotation.
These are only rough approximations but they serve to give
an indication of the possibility that rotation induced mixing may occur
in massive stars. From these estimates
we see that the circulation time is of the order of the
main sequence lifetime for rotational periods of the order of 5 days.
This corresponds to equatorial velocities of about 100 \kms\ for a star
with a radius of 10 \Rsun. The observed distribution of $v~sin~i$
of main sequence O-stars shows a bimodal distribution, with
peaks near 100 and 300 \kms\ (Conti \& Ebbetts 1977). 
This implies
that the circulation times may be of the order of the main sequence 
lifetime. Therefore we may expect rotation induced mixing to occur.
 
\subsection{The effect of rotation induced mixing on the CNO abundances}
\label{sec:6b}

In this section we describe the changes in the surface composition
due to mixing on a
{\it mean mixing timescale}, $\taumix$.
If the mixing time were short compared to the MS lifetime,
the star would be fully mixed at the end of the H-core burning phase.
In that case the CNO abundance
throughout the star would be about in CNO-equilibrium, with
the abundance ratios given by Eq. (\ref{eq:CNO-equil}) for a star of
60 \Msun.
This is obviously not the case.
If the mixing time is much longer than the MS lifetime,
rotation induced mixing would not be effective.
The most interesting case is that of partial mixing, when
\taumix\ is of the same order of magnitude as \taums.
By comparing the observed nebular abundances with
the predictions for different values of $\taumix$ we can
empirically derive the values of $\taumix$ for massive stars.

We calculate the changes in the composition of the envelope 
due to mixing by assuming that
in a time interval $\Delta t$ a fraction $\Delta t/\taumix$ of the
total mass of the star is mixed. If $X_{\rm core}(t)$ is 
the mass fraction
of an element in the convective core at time $t$ and
$X_{\rm env}(t)$ is the mass fraction of that same element in the envelope,
than the change in the mass fraction in the envelope due
to mixing can be written as
 
\begin{equation}
\label{eq:CNOmix}
\frac{d X_{\rm env}}{dt}~=~
\frac{X_{\rm core}(t)-X_{\rm env}(t)}{\taumix} \frac{M_*(t)}{M_{\rm env}(t)
}
\end{equation}
where $M_*$ and $M_{\rm env}$ are the masses of the star and of the envelope
at time $t$. This equation can be solved analytically for constant
$X_{\rm core}$ and $M_{\rm env}/M_*$.
However, the convective core shrinks, both the stellar mass
and the envelope mass change during the main sequence phase and
the composition of the core may change with time.
Therefore
we have  solved this equation numerically.
Starting with the original composition of the envelope
at the beginning of the main sequence phase 
(adopting the abundance ratios discussed in Sect. \ref{sec:4})
and taking the variation of
$M_{\rm core}(t)$, $M_{\rm env}(t)$ and $X_{\rm core}(t)$ from the
evolutionary calculations, we calculated the changes in the
surface composition due to rotation induced mixing.
In these equations we assume that the
material that is freshly mixed from the envelope into the core
is quickly transformed into the core composition
at that time. This is a good assumption
for C and N, because the CN-cycle quickly reaches equilibrium 
(see Figure 2) and
because the timescale for convective overturning in the core is very short.
But it is not a
very accurate assumption for the O-abundance in the early part of
the main sequence phase of the $Z=0.02$ stars because the ON-cycle
takes a longer time to reach equilibrium.
However since we are
mainly interested in the composition of the envelope
at the end of the main sequence phase, we can adopt
Eq. (\ref{eq:CNOmix}) as a first estimate
to describe the changes in the CNO abundances of the envelope at the
end of the core H-burning phase.
 
We note that our description of the mixing is 
a simplified version of the diffusion approximation and that 
our definition of the mixing time, defined by Eq. (\ref{eq:CNOmix}),
is not the same as the definition of the diffusion time. In the 
diffusion approximation the mixing is described as 
(Kippenhahn \& Weigert, 1990)

\begin{equation}
\label{eq:diffusion}
\frac{dX}{dt}~=~\frac{1}{dM}\left\{(4 \pi r^2
  \rho)^2~D~\frac{dX}{dM}\right\}
\end{equation}  
The diffusion coefficient is $D=l^2/\tau_D$ where $l$ is the length
scale of the diffusion and $\tau_D$ is its timescale. Using
$4 \pi r^2 \rho l \simeq M_*$ and $dM \simeq M_{\rm env}$, we find

\begin{equation}
\label{eq:diffusionmixing}
\frac{d X_{\rm env}}{dt}~=~
\frac{X_{\rm core}(t)-X_{\rm env}(t)}{\tau_D} 
\left\{\frac{M_*(t)}{M_{\rm env}(t)}\right\}^2
\end{equation}
This differs from the adopted Eq. (\ref{eq:CNOmix}) by a factor
$M_*/M_{\rm env}$.
Comparing the expression (\ref{eq:CNOmix}) and (\ref{eq:diffusionmixing})
we see that they are identical 
if $\tau_{\rm mix}=\tau_D~M_{\rm env}/M_* \simeq 0.5 \tau_D$
for massive stars.

\subsection{The effect of rotation induced mixing on the H and He abundance
}
\label{sec:6c}
 
Equation (\ref{eq:CNOmix}) can also be used to describe the H and He abundance
in the envelope, but in this case we cannot assume that all H that is
mixed from the envelope into the core will instantaneously
reach the core composition because the transformation of H into
He occurs on the main sequence time scale.
In this case the change in the H-abundance in the stellar {\it core}
is described by
 
\begin{equation}
\label{eq:hcore}
\frac{d~X_{\rm core}(t)}{dt}~=~
 -\frac{L_*(t)}{\epsilon_{H}} \frac{1}{M_{\rm core}(t)}
- \frac{X_{\rm core}(t)-X_{\rm env}(t)}{\taumix} \frac{M_*(t)}{M_{\rm core}
(t)}
\end{equation}
where $\epsilon_H$ is the energy produced by transforming
a unit mass of H into He.
The first term on the right hand side
describes the change due to H-burning
and the second term describes the changes due to mixing.
If we set $\taumix >> \tau_{\rm MS}$, we retrieve
the chemical changes in the core of the star as predicted by the stellar
evolution models without mixing. The change in the H-abundance
in the {\it envelope} 
of the star due to rotation induced mixing is described by
Eq. (\ref{eq:CNOmix}).
The changes in the He-abundances in the core and the envelope follows from
the condition that the sum of the mass fractions of H and He 
is invariant to mixing and nuclear fusion during the main sequence phase:
$X(t)+Y(t)=1-Z={\rm constant}$.
 
For calculating the variation of the surface H and He abundance
for a star with rotation induced mixing we solve the Equations
(\ref{eq:CNOmix}) and (\ref{eq:hcore}) simultaneously. We adopt the changes
 in luminosity and in the core mass and the envelope mass
as a function of time from the evolutionary
tracks of Meynet et al. (1994) with enhanced mass loss. 
For severe mixing, i.e. $\taumix < \taums$,
 this assumption is not
justified because mixing would bring fresh H into the core which would
prolong the main sequence phase. However, we will show below that the
observations indicate only mild mixing, so that the evolutionary
calculations without mixing can be adopted as a first approximation
to estimate the abundance changes due to rotation induced mixing.

We note that our description of the mixing (Eqs. 11 and 14) implies
that we predict only the  ``mean'' values of the abundances in 
the envelope as a function of time which also gives 
the evolution of the surface abundances in the 
star as a function of time.  In reality the diffusive 
rotation-induced-mixing will set up an abundance gradient in 
the envelope (see e.g. Heger et al. 2000). 
We will return to this aspect later in Sect. \ref{sec:7}.

 
\subsection{The surface abundances as a function of mixing-time}
\label{sec:6d}
 
Using Equations (\ref{eq:CNOmix}) and (\ref{eq:hcore})
we have calculated the expected surface abundances
of massive stars with rotation induced mixing. The mixing time \taumix\ is
expressed in terms of the main sequence lifetime \taums, with
$\taumix / \taums$ between 0.5 and 5.
Figure 4
shows the envelope abundance at the end of the main sequence phase,
for four initial masses.
We see that the N/O, N/C, N/S and He/H ratios 
increase as the mixing time decreases.
Even for a mixing time as long as 5~\taums ~the N/O ratio
of about unity is already significantly higher than the initial value of
N/O=1/13. This is because the N/O ratio in the core
is a factor 700 higher than the initial value. 
For such a large difference between the
initial abundance and the core abundance, the surface N/O ratio
is already increased by a factor 3 at a time as small as 
$t \simeq 0.1 \taumix$ if
$M_{\rm env} \simeq 0.5 M_*$ and the O abundance was in ON-equilibrium.
Figure 4 also shows that for strong mixing
with $\taumix \lesssim 0.5 ~\taums$ the surface abundances at the end
of the main sequence phase approach the
abundances in the core, as expected.

\begin{figure}
\label{fig:rotmix}
\centerline{\psfig{figure=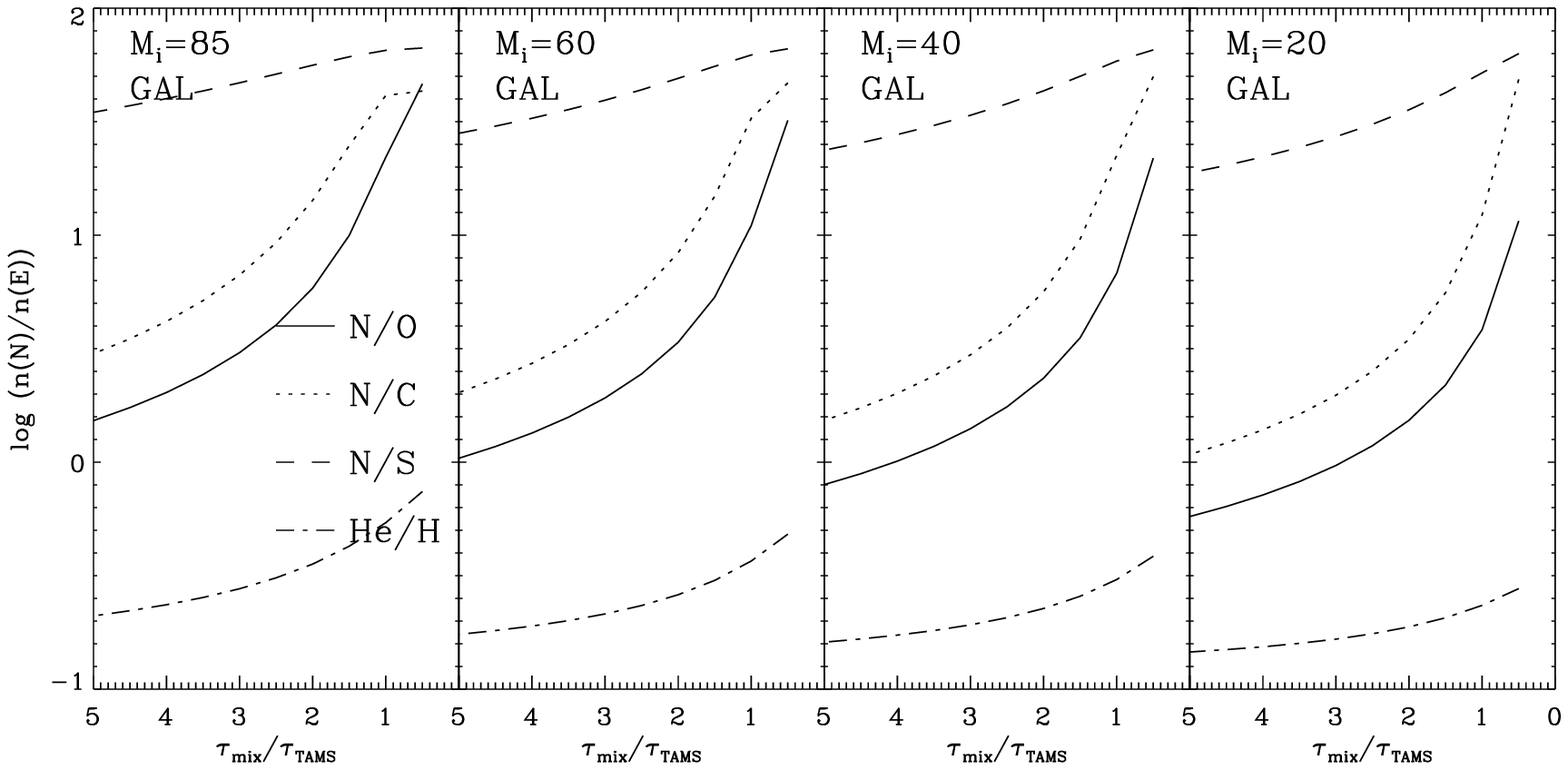,width=\columnwidth}}
\vspace{-3.5cm}
\centerline{\psfig{figure=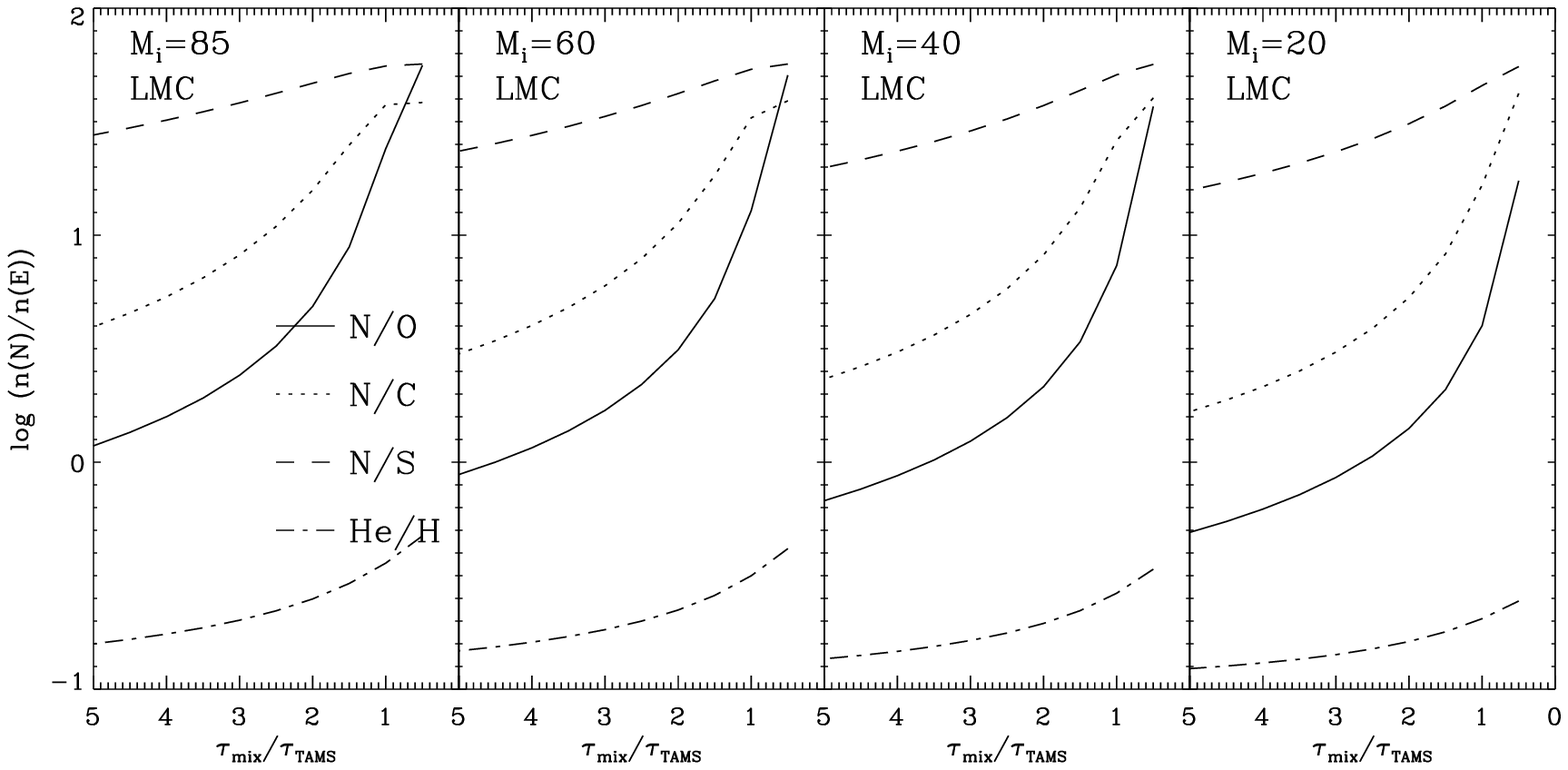,width=\columnwidth}}
\vspace{-1.0cm}
\caption[]{The logarithmic surface number ratios 
of N/O, N/C, N/S and He/H at the end 
of the main sequence phase
for stars with rotation induced mixing. The horizontal parameter
is the ratio between the mixing time and the duration of the 
main sequence phase for four initial masses.
The upper figure is for Z=0.02 and the lower figure
is for Z=0.008.} 
\end{figure}

The N/O ratio in the envelope is
smaller for stars with smaller initial masses than for high initial mass,
especially for large values of \taumixratio. This is because the
mass loss rate in the higher mass stars is larger and hence the
ratio between the mass of the envelope and the mass of the core
is smaller for high mass stars. With a smaller envelope mass,
less mixing is required to increase the envelope abundance.
Comparing the results of the $Z=0.020$ with $Z=0.008$ we find that the
envelope abundances due to mixing are very similar. The main difference
is for fast mixing which gives a higher value of N/O in the
$Z=0.008$ models than in the $Z=0.020$ models. This is because the
O abundance has a steeper profile with less O in the core of
the LMC models than in the core of the Galactic models because of the
ON-cycle reaches equilibrium composition faster for $Z=0.008$ than for
$Z=0.020$ (see Figure 2).

Heger \& Langer (2000) predict the changes in the surface abundances
of rotating stars of 8 to 25 \Msun\ using detailed stellar evolution
models. Their 20 \Msun\ model with an initial rotation velocity of
300~\kms\ evolves to $\log (N/O)=-0.05$ and $\log (N/C)=0.91$, which
compares well with our results for a star of 20 \Msun\ and 
$\taumix \simeq 1$ to 3 $\times$ $\taums$. Meynet (1998) presented models
for rotating stars of $M_i$=20 and 40 \Msun\ and finds ratios of
$\log (N/C) \simeq 0.7$ for 20 \Msun\ and  1.3 for 40 \Msun\ 
for a similar rotational velocity. This compares, again, well with 
our results for $\taumix \simeq 1$ to 3 $\times \taums$. We conclude that
our simplified prescription does yield results which are comparable 
with those of more sophisticated calculations, and that we can
use our predictions for comparison with the observations.

\subsection{Comparison with the observations}
\label{sec:6e}
 
If the LBV nebulae are ejected at the end of the main sequence phase,
but before the RSG phase, the abundance of the nebula can be used to
derive the timescale of the rotation induced mixing. The results are
summarized in Table 3. 

\begin{itemize}
 
\item{\bf AG Car:}
 If AG Car is in the post main sequence phase, its initial mass
derived from the location in the HR diagram is $M_i \simeq 70~ \Msun$.
(This mass estimate is higher than the value of $M_i \simeq 60~\Msun$ 
used in Sect. \ref{sec:5d} in case the
star has been a red supergiant, because the luminosity
of the star has increased during the RSG phase. This results in a lower
estimate of the initial mass for a given post-RSG luminosity compared
to that for a given post-MS luminosity.)
  Interpolating between the results for $M_i=60$ and $M_i=85$ \Msun\
in Figure 4 we see that the observed  ratio
N/O= $6 \pm 2$ requires
a ratio of $\taumixratio \simeq 1.8\pm 0.3$. Since the main sequence life time
of a star of 70 \Msun\ is $3.1 \times 10^6$ years (Eq. \ref{eq:taums}),
this implies a mean mixing time of $5.6 \times 10^6$ years.
For this mixing time the predicted N/S abundance is N/S = 60
and He/H = 0.33 for the nebula. The observations show that N/S $< 45$, which
is slightly smaller than the predicted value.

\item{\bf P Cygni:}
 The stellar parameters of P Cygni indicate an initial mass of
about 50 \Msun, if the star is in the post main sequence phase.
The N/S ratio of its nebula, derived from interpolation between
the models of 60 and 40 \Msun, indicates a mixing time of
$\taumixratio = 4.0 \pm 0.5$, i.e $\taumix \simeq 1.5 \times 10^7$
years. For this mixing time, the expected He/H ratio of the nebula is 0.18.
This is much smaller than the observed {\it photospheric} He/H ratio.
We return to this in Sect. \ref{sec:7}.

\item{\bf R\,127:} 
If R\,127 is a post main sequence star, its initial mass
must have been 75  \Msun.
From the interpolation between the calculations of 60 and 85 \Msun\ in
Fig. 4 we find that the abundance ratio N/O = $0.9\pm 0.4$
requires a mixing time $\taumix = 4.0 \pm 0.5 \times \taums \simeq 1.3 \times 10^7$ yrs.
For such a mixing the predicted ratio N/S is $< 30$,
which agrees with the observed upper limit for the nebula of $ <34$.
The predicted He/H ratio is 0.16, which is much smaller than the
photospheric ratio.
 
\item{\bf S\,119:}
The initial mass of S\,119 derived from its location in the
HR diagram, under the  assumption that it is a post main sequence star, is
about 65 \Msun.
From Figure 4 we see that
the observed ratio N/O= $1.9 \pm~0.5$
requires a mixing time of about $(2.9 \pm 0.4) \times ~\taums
\simeq 1.0 \times 10^7$
yrs. For such a mixing time the predicted N/S ratio is $\simeq 40$
which agrees with the observed upper limit of $<78$.
The predicted He/H ratio of the nebula is 0.18, which is much smaller
than the photospheric ratio.

\end{itemize}

We conclude that {\it if} the LBV nebula is ejected at the end
of the main sequence phase, 
mixing from the core into the envelope must have occurred
during the main sequence phase.
Using a simple mixing time description, we find that the
observed N/O ratio in AG Car, R\,127 and S\,119 and the N/S ratio of P
Cygni requires a mixing on
timescales of 1.8 to 4 times the main sequence time,
i.e. about $5 \times 10^6$ yrs for AG Car and 1.0 to 1.5 $\times 10^7$ 
years for P Cygni, R\,127 and S\,119.
We see that such relatively long mixing times can still significantly
alter the N/O and N/S ratios at the stellar surface, because only
a little mixing is necessary to change these ratios.
The fact that we find a range of two in the mixing time of stars
in such a small mass range is not surprising, as the mixing depends
not only on the initial composition and the evolution but also 
on the initial rotation velocities. These might differ from star to star.

We can
estimate the rotation velocity during the zero age main sequence phase
that is needed to explain the derived mixing timescale. The result is listed
in Table 3 
where we have used Eqs. (\ref{eq:mixingtimegal}) and
(\ref{eq:mixingtimelmc}) and the ratio $\tau_{\rm circ} \simeq
\tau_D \simeq 0.5~\tau_{\rm mix}$. We see that the required 
equatorial rotation velocities are in the range of 70 to 100 \kms.
These values are not particularly high for massive main sequence
stars. Conti \& Ebbets (1977) found that the distribution in $v$~sin~$i$
of main sequence O-stars has a bimodal distribution with peaks at
100 and 300 \kms. So it is possible 
that the enhanced N/O ratios observed in LBV nebulae already 
appear at the surface of the stars near the end of the main sequence
phase due to rotational mixing. 
In fact the presence of the ON-supergiants is a clear indication
of this process (Walborn, 1988; K.C. Smith et al. 1998), as well as the existence
of N enriched main sequence stars in the LMC (Korn et al. 2000).
Unfortunately no detailed quantitative studies of the C, N and O abundances
of the ON stars has been done yet, to see if the
abundances are the same as in LBV nebulae.


\section{When were the LBV-nebula ejected?}
\label{sec:7}

We have shown above that the abundances of the LBV nebulae, in particular 
the N/O ratio can be explained by ejection in either the RSG phase,
when mixing in the convective envelope has occurred, or 
at the end of the main sequence phase, hereafter referred to as the 
blue supergiant (BSG) phase, if rotation induced mixing occurred
during the main sequence phase.
In this section we compare the results derived above for the 
two explanations with the 
constraints of the dynamical age of the nebula and the
present photospheric He/H ratio (see Tables 1, 2 and 3).
These constraints are:

\begin{enumerate}

\item The dynamical ages of the LBV nebulae are between 1 and 7 times
$10^4$ years, apart from the P Cygni nebula which is much younger.

\item The LBV nebulae contain between about 2 and 8  solar masses
of mildly enriched ionized gas, except for P Cygni which has a much
smaller nebular mass.
The amount of molecular or neutral material in LBV nebulae is not well known. 
A recent study of the CO emission from the AG Car nebula 
(Nota et al. 2000b) shows that the nebula contains about
$5 \times 10^{-3}$ \Msun\ in the form of CO. The fraction of CO 
is estimated to be $\sim 2.3 \times 10^{-3}$ of the total
molecular gas, which then amounts to 2.6 \Msun. This is about
as much as the 4.2 \Msun\ of ionized gas in the AG Car nebulae.
So the total amount of mass in LBV nebulae might be as high as
twice the mass of the ionized gas.

\item The photosphere of the LBVs is more chemically enriched 
than the nebula.
The N/O ratio in the nebulae suggest a He/H ratio of about 0.3 for 
the nebula of AG Car and about 0.2 for the nebulae around P Cygni, 
R\,127 and S\,119 (Sections \ref{sec:5d} and \ref{sec:6e}).
However, the photospheric He/H ratios are between 0.4 and 0.7.

\item The outflow velocities of the LBV nebulae are 
70 \kms\ for AG Car and 140 \kms\ for P Cyg. The outflow velocities
of the LMC stars R\,127 and S\,119 are 25 and 29 \kms\ respectively.

\item No red supergiants with luminosities above 
$6~10^5$ \Lsun\ have been found in the Galaxy or in the LMC (Humphreys
\& Davidson 1979).
This implies that either more luminous stars do not pass through a RSG phase
or that their RSG phase lasts only a very short time, i.e. less than
about a few $10^4$ years.   

\end{enumerate}

\subsection{The ejection of the nebula at the beginning of the RSG
  phase?}
\label{sub:beginningrsg}

 Let us suppose that the nebula was ejected during the RSG phase. 
From the nebular abundance we found that AG Car must have lost
about 25 \%\ of its initial mass prior to the onset of the envelope convection.
The convection then created a layer of about 15 \Msun\ of mildly
enriched material above the non-convective core. 
If the observed $\sim$ 7 \Msun\ of the nebula (ionized plus molecular gas) 
was ejected near the beginning of the 
RSG phase, i.e. before significant RSG mass loss has occurred, 
the star would still have an envelope of about 8 \Msun\
of mildly enhanced material after the ejection.  This is at odds with 
the fact that about $10^4$ years after the nebula was ejected 
the He/H ratio at the surface of the star has already increased 
significantly to the observed value of He/H $\simeq$ 0.42.
It would require an unrealistically high mass loss rate of about
$10^{-3}$ \Msunyr\ after the ejection of the nebula.
The problem is even more severe for the LMC stars which have lost
a smaller fraction of their initial mass before the envelope convection
starts. They will have a much more massive layer 
with mildly
enriched material and a predicted He/H ratio of only about 0.20.
In order for these stars to reach a ratio He/H $\simeq$ 0.4 at the
surface requires the ejection of even more mass than from the Galactic LBVs
within the short dynamical age  of their nebula.

\subsection{The ejection of the nebula at the end of the RSG phase?}

If the nebula was ejected near the {\it end}
 of the RSG phase, when the star had
already lost about 10 \Msun\ as a RSG, the increase in surface abundance to 
the presently observed He/H ratio of about 
0.4 within the dynamical age of the nebula
can  be explained because soon afterwards the deeper more enriched layers
may have been exposed. Adopting a ``normal'' RSG mass loss rate
of about $10^{-4}$ \Msunyr, the star may have lost its last
solar mass of mildly enriched material in the dynamical time
of the nebula. 

 The question is then, where is the mass 
that must have been ejected in the RSG phase {\it before} the nebula was
ejected? 
For example, AG Car had about 15 \Msun\ of convectively mixed material
in the early-RSG phase, about 7 of which are in the nebula 
($\sim$ 4 \Msun\ ionized and $\sim$ 3 \Msun\ molecular). The other 8 \Msun\
must have been lost before the nebula ejection.
With a typical RSG wind velocity of 10 \kms\ that material
should have reached a distance of about a few tenth of a parsec, and
it would have been overtaken by the nebula that was ejected with a velocity
of about 70 \kms\ for AG Car. So in that case we would have expected
the visible nebula to contain practically all the gas that 
was lost during the RSG phase, i.e. about 15 \Msun. This is not observed.
Moreover, in that case we cannot explain 
the lack of luminous RSG, because they must have lived about $10^5$ years
to eject the initial 10 \Msun\ layer of the mixed material (unless
the mass loss rate was very high during that phase).

\subsection{The ejection of the nebula in the BSG phase ?}

Let us now suppose that the nebula was ejected in the BSG phase,
and that rotation induced mixing has affected the abundances
as predicted in our calculations.
(These abundances are confirmed for stars in the mass
range 15 of 25~\Msun\ by the detailed evolutionary models for rotating stars
by Heger (1998, p 81) and Heger et al. (2000).)
In this case, the N/O and N/S ratios can all be
satisfactorily explained, although this requires differences in the
adopted mixing time scales in the stars by about a factor of two.
 
The increase of the CNO-enrichment from
the mild enhancement in the nebula, which indicates  He/H $\simeq$ 0.2,
to the present day photospheric
He/H ratio of about 0.4 to 0.7 can also be explained in the framework of
the rotating stellar models by two effects which will both be
operating:\\
(1) Rotation induced mixing will continue after the ejection of the nebula
in the BSG phase and the He/H ratio will keep increasing with time.
In fact, the mixing is likely to become faster when the star
moves away from the main sequence, because it may approach its
$\Omega$ limit for near-critical rotation in the layers where the
radiation pressure is high (Langer 1998; Langer et al. 1999).
For near-critical rotation
the mixing-time approaches the Kelvin-Helmholtz time, which is about
$10^4$ years for a star with an initial mass of 60 \Msun. This means that
within about $10^4$ years after the ejection of the nebula the 
composition may have changed drastically.\\
(2) Rotation-induced-mixing is a diffusive process that will result
in an abundance gradient in the envelope, with the most enriched
matter closer to the core (see Sect. \ref{sec:6c}). The ejection of the
outer layers will therefore automatically reveal photospheric layers
which are more enriched than the ejected gas.

In this respect the star P Cygni is an interesting test case. The
nebula was ejected only 2000 years ago. We do not know the He/H ratio
in the nebula, but the N/S ratio implies that the He/H ratio in the 
nebula must be about 0.2, whereas the He/H in the photosphere is 0.4.
\footnote{Unfortunately, the abundance determination of the P Cygni
nebula is less well known than that of the other program stars. A
more accurate determination of the abundance and of the 
total amount of ionized plus neutral gas would be highly desirable.}
Such a difference implies that there must have been an abundance
gradient near the outer layers of the star. An abundance gradient
is expected for rotation induced mixing by diffusion 
but not
for mixing by convection in the RSG phase. Our simplified 
method of predicting the rotation-induced-mixing does not 
allow us to calculate the abundance gradients.
Heger et al. (2000) have calculated the evolution of rotating stars
in the mass range of 10 to 25 \Msun\ with diffusive mixing.
Although these models have lower masses than P Cygni, they 
indicate that abundance gradients of the required order
may well be present in rotating massive blue supergiants.

Additional evidence for a prominent role of rotation
in the ejection of the nebula comes from the observed
bipolar structure of several LBV nebulae (Nota, 1997)
and the linear polarization of the UV and optical radiation
of AG~Car and R~127 (Schulte-Ladbeck et al. 1994).

A more indirect argument for the ejection of the LBV nebulae in the
blue supergiant  phase may come from the expansion velocities 
of the nebulae. The expansion velocity of the Galactic nebulae of AG Car and 
P Cygni are 70 and 140 \kms\ respectively. The wind
speeds of luminous red supergiants are smaller and on the order of 10 to 20
\kms\ (Reimers 1975, Dupree \& Reimers, 1987).
Although the ejection of the nebulae may have been due to a process
other than the normal RSG wind, it is unlikely that this process could
have lead to an ejection velocity much higher than that of the wind.
In general, an {\it increase} in the mass loss rate by some 
subphotospheric mechanism results in a {\it decrease} of the velocity
(unless a large amount of extra energy is generated as in Nova 
outbursts). 
On the other hand, the wind velocities of luminous 
blue supergiants with high mass loss rates such as 
$\zeta^1$ Sco or $\alpha$ Cyg are about 100 to 200 \kms. 
An increase in the mass loss rate by a significant factor,
such as during the ejection of the nebula, will result in a decrease of 
the wind velocity. So it is easier to explain the velocities
of the nebulae of P Cyg and AG Car if they are ejected in the BSG
phase than in the RSG phase. 
(The velocities of the nebulae of the LMC stars R\,127 (29 \kms) 
and S\,119 (25 \kms) are markedly smaller than those of the Galactic 
counter parts. In fact they are close to those expected for red supergiants.
Until we understand the ejection mechanism of the LBV nebulae
and its dependence on metallicity, it is difficult to explain the
reason for the difference in nebular velocity between the Galactic
LBVs and their counterparts in the  LMC.)


\section{Discussion and conclusions}
\label{sec:8}

In an attempt to explain the origin of the LBV nebulae, 
which are mildly enriched in N and depleted in O, we have
studied the effects that can modify their composition.
The results are summarized.

\begin{itemize}

\item{} We have calculated the expected abundance changes at the
stellar surface due to two effects: (a) envelope convection
in the RSG phase (Sect. \ref{sec:5}) and (b) rotation induced mixing
on the main sequence (Sect. \ref{sec:6}). 

\item{} We have shown that the abundance patterns of four LBV nebulae,
in particular the N/O and N/S ratios, can be explained by both
mechanisms:\\
-- either, by mixing due to an outer convective envelope. In that case
the LBV nebula must have been ejected during the red supergiant
phase. The observed N/O and N/S ratios then provide an estimate of the
amount of mass that must have been lost before the envelope convection
sets in. For this scenario, we find that the mass lost 
is significantly smaller than adopted
in the evolutionary calculations of Meynet et al. (1994).\\
-- or, due to rotation induced mixing during the main sequence phase.
In that case the LBV nebula was ejected after the main sequence
in the blue supergiant phase. The N/O and N/S ratios then provide an
estimate of the mixing time. The mixing times are between
$5 \times 10^6$ and $1.5 \times 10^7$ years. These values are
reasonable, considering the fast rotation of the main sequence
O-stars. Additional evidence that rotational mixing is common in
early type stars comes from the fact that a considerable 
fraction of the OBA supergiants and even some O main sequence stars 
have an enhanced N abundance (e.g. Walborn, 1988; 
K.C. Smith et al. 1998; Venn 1996, 1997, 1999)

\item{} Both models (convective or rotation induced mixing)
predict a nebular He/H ratio that is about a
  factor two smaller than the observed {\it photospheric}
abundance of He/H $\simeq$ 0.4 to 0.7. This enrichment of the stellar 
surface must have occurred after the nebula was ejected, i.e. 
within the dynamical timescale of the nebulae of about $10^4$ years.
This cannot be explained in the context of the nebular ejection
during the RSG phase, but it is a logical consequence of  
rotation induced mixing for two reasons:\\
(1)  The mixing is expected to become faster as the star 
expands after the main sequence phase and approaches its $\Omega$
limit for critical rotation in the layers where the radiation pressure is
high. 
Depending on their Eddington-factors, 
for stars near critical rotation the mixing time may become as
short as 
the Kelvin-Helmholtz time, which is of order $10^4$ years for a star
of 60 \Msun.\\
(2) Contrary to convective envelope mixing, rotation-induced-mixing 
is a diffusive process that results in an abundance gradient in 
the envelope with the more enriched material  closer to the core.
The ejection of surface layers will automatically reveal the more
enriched deeper layers.

\item{} The expansion velocities of about 100 \kms\ of the 
LBV nebulae of AG~Car, P Cyg and R~143 can be explained
if the nebulae were ejected during the BSG phase but not during the
RSG phase. The low expansion velocities of R~127 and S~119 might be the
result of deceleration by interaction (these are the oldest nebulae
of our sample) or due to the small radiative acceleration because of the low
metallicity of the LMC stars.

\end{itemize}

We conclude that the abundance pattern observed in LBV nebulae and in the
photospheres of their central stars may be explained
with the rotation induced mixing scenario. This scenario may explain 
all observations
consistently. In the rotation scenario the differences in the N/O
ratios of the LBV nebulae are due to different mixing time scales,
that are most likely due to differences in the initial rotation rates.
An advantage of the mixing scenario
may be that it ties in with the LBV outburst model involving the
$\Omega$-limit (Langer et al. 1999),
which appears to be able to explain the bi-polar morphology
found in virtually all LBV nebulae (Nota et al. 1995).
It would be interesting to investigate if the model for the 
outburst due to the $\Omega$-limit might also
explain the fact that the photospheric He/H ratio of the LBVs is in
a narrow range of 0.4 to 0.7. This 
narrow range might possibly be due to the fact that in such a model, 
the rotation determines both the moment of the outburst and the
amount of mixing.

How do we combine the scenario that the LBV nebulae are ejected in the
BSG phase with the results of the studies of the ISO
spectra of these nebulae by Waters et al. (1997; 1998; 1999) and Voors et al.
(2000). These authors have shown that the dust in the LBV nebulae
is very similar to the dust around red supergiants, i.e. 
mainly in the form of
amorphous silicates plus a minor contribution from
crystalline silicates such as olivines and pyroxenes.
Since the composition depends sensitively on the conditions during
the ejection of the material, Waters et al. (1997; 1998; 1999) have argued 
that the LBV dust must have been ejected when the star was a RSG.
However, one has to be careful with applying this argument and one should
make a distinction between two possible interpretations: \\
(a) The star was a red supergiant in the evolutionary sense,
i.e. it had  a large convective envelope during a typical evolutionary
timescale of order $10^4$ to $10^5$ years. (We have rejected this
possibility on the basis of the abundances and velocities of the
LBV nebulae)
\\
(b) During the ejection of the nebula, by whatever mechanism,
the star was temporarily cool and large in size, resembling a 
RSG. This is likely  because a very high temporary mass loss rate
will produce an extended cool pseudo-photosphere because of its 
large optical depth. The dust that forms outside such a photosphere
will be formed under similar conditions as in RSG winds. This
explanation agrees with the arguments that the nebula was
ejected in a {\it temporary} RSG phase.

We conclude that many  arguments point to the LBV nebulae
being ejected during the blue supergiant phase and that the
chemical enrichment is due to rotation induced mixing.

 This suggests the following {\it evolutionary scenario} for  the
formation of the LBV nebulae with the observed chemical
and dynamical properties:\\
1. Mildly or rapidly rotating massive stars experience rotation
induced mixing during the MS phase. This enhances the He/H ratio
from the initial value of 0.1 to about 0.2 to 0.4 and the N/O ratio from 
0.07 to about 1 to 5 in the envelope, depending on the
ratio between the mixing time and main sequence life time. 
This explains the existence of the ON-stars.\\
2. After leaving the main sequence and evolving into
blue supergiants, the stars run into the $\Omega$-limit,
which is the critical rotation limit in a star with
a significant contribution by radiation pressure.\\
3. When the star reaches the $\Omega$-limit it ejects a large
amount of mass, on the order of one or a few solar masses. The 
ejected matter has an enhanced N/O ratio, because it consists of 
matter from the
envelope that was mixed during the main sequence phase.
This explains the observed abundances of LBV nebulae.\\
4. When the star has reached
thermal equilibrium after the ejection, 
it is again located to the left of the $\Omega$-limit
in the HR-diagram.  It will then start again to expand 
and will reach  the $\Omega$-limit. In this way 
the star may suffer multiple ejections, with time intervals
on the order of a fraction of the Kelvin-Helmholtz timescale, i.e. about
$10^3$ to $10^4$ years. This may explain the relatively large number of
eruptions in Galactic LBVs (two in about 6 LBVs) 
that have been observed during the last four hundred 
years. \\
5. If the ejections of a large amount of mass are  due to the
star reaching its $\Omega$-limit, we can expect that the mass
is ejected mainly in the equatorial plane. 
(Models of moderately rotating stars show that the mass flux
from the poles is slightly larger than from the equator, e.g. Maeder (1999).
This will not be the case for a star near critical rotation. In any
case, the density of the equatorial wind will always be higher than
the density of the polar wind, because the wind velocity is very
sensitive to the local escape velocity and will therefore be lower
in the equatorial region than in the polar region, 
 e.g. Pelupessy et al. (1999).)
The nebula ejected at the first ejection (when the star has moved off
the main sequence) will be influenced very little 
by the wind form the main sequence phase, because the
main sequence wind velocity is expected to be higher
than the velocity of the ejecta. However, the nebula 
formed by the later ejecta
may be shaped by the presence of the equatorial
circumstellar matter from the previous ejection or from the 
equatorially enhanced winds from the rapidly rotating star
in between the ejections. This may explain the observed bipolar structure of
most LBV nebulae.\\
6. During the ejection phase, the outflow is optically thick, which 
results in a large effective radius and a low effective temperature
of the star that temporarily resembles a red supergiant. 
This explains the similarity between 
the properties of dust in LBV nebulae and in RSG winds.\\
7. After each ejection, the He/H ratio and the N/O ratio
of the stellar photosphere is (slightly) higher than that of the
ejected gas, because there is an abundance gradient in the 
envelope due to the continued rotation induced mixing of the envelope
with the deeper layers. The mixing gets faster and the abundance 
gradient gets steeper, the closer the star moves to the 
$\Omega$-limit. This may explain why the photospheric He/H ratio
is higher in the LBV photospheres than in their nebulae.\\ 
8. The ejections stop when the star has lost most of its
H-rich envelope. The star then contracts to become a Wolf-Rayet
star with a He-rich envelope.\\

\section{Acknowledgement}
H.J.G.L.M.L., L.J.S. and N.L. are 
grateful to the Space Telescope Science Institute
for financial support and hospitality. We thank Robert Voors and
Rens Waters for enlightening discussions on the dust properties
of LBV nebulae, and Nolan Walborn
for discussions about chemically enriched ON stars.
We thank George Meynet for critical comments and suggestions
on an earlier version of this paper.
This work has been supported by the Deutsche Forschungsgemeinschaft
through grants La~587/15-1 and 16-1 to N.L.


\clearpage
\begin{table}
\caption{Stellar and nebular parameters for LBVs}
\footnote{Crowther and Smith 1997 found lower L for S\,119. check!}
\label{tbl:parameters}
\begin{tabular}{lccccccc}
\hline
\hline
 Name       & $d$          & log \Lstar    & He/H\tablenotemark{a}   
 & $M_{\rm neb}$ & $v_{\rm exp}$& $\tau_{\rm dyn}$
 &    \\ 
            & kpc          &    \Lsun      & phot          &  \Msun &
 \kms       &  yrs     &    \\
\hline
 AG Car     &  6           & 6.0           & 0.42          &  4.2 &
    70      &  $1\times 10^4$ &   \\
 P Cyg      &  1.8         & 5.8           & 0.40          &  0.01 &
    140     &  $2.1\times 10^3$     &   \\
 R\,127       & 51.2         & 6.1           & 0.50          &  7.9  &
    29      & 2 \& 7x$10^4$   &   \\
 S\,119       & 51.2         & 6.0           & 0.67          &  2.5  &
    25      &  5 $10^4$       &   \\
\hline
\hline
Name & \nel & \Tel          & [N/H]\tablenotemark{b}      & [O/H]\tablenotemark{b}
&  N/O\tablenotemark{a}      & N/S\tablenotemark{a}  & (N+O)/H\tablenotemark{a}   \\
   &  cm$^{-3}$     & K       &       &       &      &        &      \\
\hline
 AG Car       & 820$\pm$170  & 6350$\pm$400 & 0.7$\pm$0.1 &
 $-1.2\pm$0.2 & $6 \pm 2$    & $ < 45$      & $4 \times 10^{-4}$ \\
 P Cyg        & 600          & 5300         & 0.75        & 
 $<0.43$      & $>0.16$      & 33           & 0.4 to
 $3\times 10^{-3}$\\
 R\,127         & 720$\pm$90   & 6400$\pm$300 & 1.0$\pm$0.1 & 
 $-0.3\pm$0.2 & 0.9$\pm$0.4  & $<34$        & $3.2 \times 10^{-4}$\\
 S\,119         & 680$\pm$170  & 6200$\pm$600 & 1.2$\pm$0.2 & 
 $-0.4\pm$0.3 & 1.9$\pm$0.5  & $<78$        & $3.8 \times 10^{-4}$\\
\hline
\end{tabular}
\tablenotetext{a}
{The ratios He/H, N/O, N/S and (N+O)/H are by number.}
\tablenotetext{b}
{The logarithmic over- and underabundance ratios are relative to the
abundances of H\,II regions in the Galaxy
(for AG Car and P Cyg) or the LMC (for R\,127 and S\,119).}
\end{table}

\clearpage
\begin{table}
\label{tbl:RSGresults}
\caption[]{The LBV nebulae ejected during the RSG phase}
\begin{tabular}{lllcclll}
\hline
\hline
Name & $M_i$ & Ratio & $M_{RSG}/M_i$ & $(\Delta M)_{\rm abund}$ & 
$(\Delta M)_{\rm evol}$ & He/H  &  He/H \\
     & \Msun\ &      &               & \Msun\                   & 
\Msun\                  & pred. &  star \\
\hline\\
AG Car & 60 & N/O & 0.78$\pm$0.03 & 13$\pm$2    & 21 & 0.32 & 0.42 \\
P Cygni& 50 & N/S & 0.93$\pm$0.03 & 3.5$\pm$1.5 & 15 & 0.22 & 0.40 \\
R\,127   & 75 & N/O & 1.00          & 0           & 25 & 0.10 & 0.50 \\
S\,119   & 60 & N/O & 0.93$\pm$0.05 & 4.2$\pm$3.0 & 15 & 0.20 & 0.67 \\
\hline
\end{tabular}
\end{table}

\begin{table}
\label{tbl:mixresults}
\caption[]{The LBV nebular abundances due to rotation induced mixing}
\begin{tabular}{llllllcll}
\hline
\hline
Name & $M_i$ & Ratio & $\taumixratio$ & $\tau_{\rm mix}$ & 
  P  &  $v_{\rm rot}^a$ & He/H  & He/H \\
     & \Msun\ &      &               &  yrs              &
 days&  \kms\           & pred. & star \\
\hline\\
AG Car & 70 & N/O & $1.8 \pm 0.3$ & $5.6\times 10^6$ & 6 & 100& 0.33 &0.42\\
P Cygni& 50 & N/S & $4.0 \pm 0.5$ & $1.5\times 10^7$ & 8 &  70& 0.18 &0.40\\
R\,127   & 75 & N/O & $4.0 \pm 0.5$ & $1.3\times 10^7$ & 9 &  90& 0.16 &0.50\\
S\,119   & 65 & N/O & $2.9 \pm 0.4$ & $1.0\times 10^7$ & 8 & 100& 0.18&0.67\\
\hline
\end{tabular}

$^a$ These are the rotation speeds on the main sequence
that are required to produce the observed nebular abundances
at the surface of the star by rotation induced  mixing.
\end{table}

\end{document}